\documentclass[preprint,10pt,sort&compress,3p,onecolumn]{elsarticle}

\usepackage{booktabs}
\usepackage{array}
\usepackage{paralist}
\usepackage{verbatim}
\usepackage{helvet}
\usepackage{amssymb}
\usepackage{mathrsfs}
\usepackage{graphicx}
\usepackage{subfigure}
\usepackage[sumlimits]{amsmath}
\usepackage{geometry}
\usepackage{fancyhdr}
\usepackage{sectsty}
\allsectionsfont{\sffamily\mdseries\upshape}

\journal{Journal of Magnetic Resonance}

\begin{document}

\begin{frontmatter}

\title{Application of Optimal Control to CPMG Refocusing Pulse Design}


\author[mit,iqc]{Troy W. Borneman\corref{cor1}\fnref{fn1}}
\ead{troyb@mit.edu}

\author[sdr]{Martin D. H\"urlimann}

\author[mit,iqc,pi,uwchem]{David G. Cory}

\cortext[cor1]{Corresponding author.}
\fntext[fn1]{Present address: Institute for Quantum Computing, 200 University Ave. W., Waterloo, ON, Canada N2L 3G1}

\address[mit]{Department of Nuclear Science and Engineering, Massachusetts Institute of Technology\\
                  Cambridge, MA, USA}
\address[iqc]{Institute for Quantum Computing, Waterloo, ON, Canada}
\address[pi]{Perimeter Institute for Theoretical Physics, Waterloo, ON, Canada}
\address[sdr]{Schlumberger-Doll Research, Cambridge, MA, USA}
\address[uwchem]{Department of Chemistry, University of Waterloo, Waterloo, ON, Canada}
                 
\begin{abstract}
We apply optimal control theory (OCT) to the design of refocusing pulses suitable for the CPMG sequence that are robust over a wide range of $B_0$ and $B_1$ offsets. We also introduce a model, based on recent progress in the analysis of unitary dynamics in the field of quantum information processing (QIP), that describes the multiple refocusing dynamics of the CPMG sequence as a dephasing Pauli channel. This model provides a compact characterization of the consequences and severity of residual pulse errors. We illustrate the methods by considering a specific example of designing and analyzing broadband OCT refocusing pulses of length $10 t_{180}$ that are constrained by the maximum instantaneous pulse power. We show that with this refocusing pulse, the CPMG sequence can refocus over 98$\%$ of magnetization for resonance offsets up to 3.2 times the maximum RF amplitude, even in the presence of $\pm 10\%$ RF inhomogeneity.

\end{abstract}

\begin{keyword}
CPMG \sep Broadband refocusing pulses \sep Dynamical decoupling \sep Optimal control theory \sep Quantum information theory


\end{keyword}

\end{frontmatter}

\section{Introduction}
\label{sec:Introduction}
A common aspect of the continuing development of magnetic resonance and quantum information processing (QIP) is the study of the efficiency, capabilities, and limits of control over complex quantum dynamics. To this end, optimal control theory (OCT) \cite{Pontryagin:62a} is an important tool in both fields for the design of precise control sequences. From the standpoint of QIP, OCT has emerged as the best method of optimizing arbitrary unitary dynamics, allowing magnetic resonance to continue serving as the most successful testbed for the development and evaluation of techniques to accurately control quantum information \cite{RyanLaflamme:08a,HavelCory:02a,FortunatoCory:02a,PraviaCory:03a}. The relevance of OCT to magnetic resonance has also been well-established \cite{ConnollyMacovski:86a,MaoAndrew:86a,RosenfeldZur:96a,GlaserGriesinger:98a,KhanejaGlaser:05a}. Here we focus on a specific instance of that relevance by considering how QIP and OCT may be applied to the Carr-Purcell-Meiboom-Gill (CPMG) \cite{CarrPurcell:54a,MeiboomGill:58a} sequence. The CPMG sequence is an ideal candidate to investigate the capabilities of OCT in both the QIP and magnetic resonance settings due to its importance to both fields and its well known design and performance criteria.

The CPMG sequence is primarily used in magnetic resonance and QIP for situations where significant field inhomogeneities - which may or may not depend on time - are present. The versatility of the sequence lies in its inherent robustness to pulse nutation angle errors and insensitivity to variations in the static, $B_0$, and the applied radio-frequency (RF), $B_1$, fields. In the context of magnetic resonance, the CPMG sequence is most useful to monitor dynamic processes, such as relaxation and diffusion. For complex systems with multi-exponential decays, it is necessary to acquire large numbers of echoes with short echo spacings to cover the entire range of relaxation times. Averaging over the multiple echoes generated by the sequence further enhances the signal-to-noise ratio (SNR) of these measurements. In one-sided and stray field magnetic resonance applications, where the inhomogeneity in $B_0$ is typically much larger than the available RF amplitude, the CPMG sequence is particularly important. These applications include well-logging \cite{GoelmanPrammer:95a,HurlimannGriffin:00a}, the NMR-MOUSE \cite{BalibanuBlumich:00a}, and ex situ and single-sided NMR \cite{MerilesPines:01a,PerloBlumich:05a}. The CPMG sequence has also proven to be the best means of suppressing general environmental decoherence in quantum computations, towards the goal of fault-tolerance \cite{ViolaLloyd:98a,BiercukBollinger:09a,CappellaroCory:06a,KhodjastehLidar:07a}. Although the CPMG sequence is inherently tolerant of field inhomogeneities and pulse nutation angle errors, the robustness of the sequence is ultimately limited by the quality of the refocusing pulses. Square hard pulses, for example, only effectively refocus spins at static $B_0$ fields that deviate from the Larmor condition by much less than the RF amplitude of the pulse. When RF field inhomogeneity is taken into account, the refocusing properties of the sequence are further degraded \cite{HurlimannGriffin:00a}. 

In almost all circumstances, it is important to develop refocusing pulses for the CPMG sequence that simultaneously have a large bandwidth with respect to static field inhomogeneity, account for RF inhomogeneity, and act as a universal rotation - as opposed to the state to state transfer performed by inversion pulses. The specific demands of each application, however, may vary. For example, in magnetic resonance measurements, maximixing SNR is often the most important criteria for designing CPMG sequences. This entails - assuming a limit on instantaneous or overall power consumption - finding the tradeoff between maximixing the bandwidth of the sequence with respect to inhomogeneity while minimizing the pulse length to allow the acquisition of more echoes in a given time period. On the other hand, for QIP applications, the pulse length is less important than maximizing the bandwidth of the sequence in order to suppress as much decoherence as possible. In order to address SNR concerns, we must first be able to systematically maximize bandwidth for a given pulse time. It is this problem that we consider in detail in this work. It directly addresses the concerns of QIP, demonstrates and evaluates our techniques, and lays the groundwork for future studies.

Finding a global solution to maximizing bandwidth for a given pulse time and subject to constraints of pulse power has proven to be difficult \cite{Freeman:98a}, and perhaps impossible due to the complexity and size of the problem. However, by taking advantage of extensive theoretical and numerical results in quantum control, we may gain insight into the problem. In this work we describe how OCT provides a systematic means of finding consistent, effective solutions to the problem of designing broadband refocusing pulses suitable for use in the CPMG sequence. We focus, in particular, on finding refocusing pulses that perform an identical unitary operation over as large a range of resonance offsets and RF amplitudes as possible, while adhering to constraints of pulse length and maximum instantaneous RF power. We also describe some future applications for which the techniques outlined in this work may be used to find novel pulses, and introduce a new method, based on the study of quantum channels from quantum information theory, to compactly analyze the dynamics and accumulation of pulse errors in the CPMG sequence.



\section{CPMG Criteria}
\label{sec:cpmgcriteria}
The CPMG sequence consists of periodic applications of a cycle, [$\tau$ - $\pi)_y$ - $2\tau$ - $\pi)_y$ - $\tau$], which repeatedly refocuses transverse spin magnetization, leading to a train of echoes. The accumulation of pulse errors during the sequence causes only a single orthogonal component of the input state to be preserved - the component along the axis of the refocusing pulses. Meiboom and Gill's modification to the original Carr-Purcell sequence was to recognize this symmetry of the pulse errors, and shift the phase of the excitation pulse applied before the sequence such that the initial spin magnetization aligns with the axis of refocusing. There are, then, two common measures for the success of a CPMG sequence: (1) a single orthogonal component of the input state does not decay in the absence of relaxation, regardless of the time between refocusing pulses (echo spacing) and (2) that the echo visibility be maximal. These measures dictate the design requirements for the refocusing pulses: (i) the direction of the effective rotation axis must be oriented exactly in the $xy$-plane, (ii) the effective rotation axis and the initial spin magnetization must be aligned, and (iii) the effective nutation angle must be $\pi$ radians.

The influence of not keeping the effective field direction in the transverse plane is seen by considering the efficiency of averaging static field inhomogeneities by a single hard $\pi$-pulse that is tilted an angle $\zeta$ from the $xy$-plane. The zeroth order contribution to the average Hamiltonian of a magnetic field inhomogeneity, $H_{\mathrm{int}} = \Delta\omega\sigma_{z}$, for the sequence $\tau - \pi)_y - \tau$ is

\begin{equation}
  \bar{H}^{\mathrm{CP}}_{\mathrm{eff}} = \Delta\omega\sigma_{z} (1 - \cos2\zeta) + \Delta\omega\sigma_{y} \sin2\zeta.
\end{equation}
Ideally, $\zeta = 0$ and the Hamiltonian vanishes. Notice that when the RF field is tilted from the transverse plane the refocusing under this sequence is incomplete and similar to that engineered in to the Chemical Shift Concertina sequence \cite{EllettWaugh:69a}, which was designed to scale chemical shifts. In calculating (1) we have kept the nutation angle fixed to $\pi$ as we wish to explicitly illustrate the decrease in averaging as the effective field is rotated out of the $xy$-plane. The residual contribution to $\bar{H}^{\mathrm{CP}}_{\mathrm{eff}}$ arises from incomplete modulation of the spin dynamics. This reduction in the modulation depth of the averaging is also detrimental to effective decoupling as described by Waugh \cite{Waugh:82a}.

Next, consider the sensitivity of the CP cycle to a mis-setting of the pulse nutation angle. We can observe this dependence by calculating the propagator corresponding to one cycle of the CP(MG) sequence, with pulse nutation error $\epsilon$, applied to a spin whose Larmor precession is off-resonance by an amount $\Delta\omega$:

\begin{equation}\begin{split}
  U(\epsilon,\Delta\omega) = & e^{-i\frac{1}{2} \Delta\omega\sigma_{z}\tau} e^{-i\frac{1}{2}(\pi-\epsilon)\sigma_{y}} e^{-i\Delta\omega\sigma_{z}\tau} \\& \times e^{-i\frac{1}{2}(\pi-\epsilon)\sigma_{y}} e^{-i\frac{1}{2}\Delta\omega\sigma_{z}\tau}.
\end{split}\end{equation}
For the sake of simplicity, the pulses are assumed to be on-resonance. The improvement that Meiboom and Gill brought to the Carr-Purcell cycle was to recognize that the refocusing is more robust when the pulse rotation axis and the initial magnetization are aligned. The benefit of the CPMG sequence is seen by examining the retained magnetization under conditions of maximal alignment and minimal alignment - $\rho_{\mathrm{in}} = \sigma_{y},\sigma_{x}$ respectively. The overlap of initial and final magnetization under the action of the cycle is

\begin{equation}
  O_{x,y} = \frac{1}{2} \mathrm{Tr}\left\{\sigma_{x,y} U(\epsilon,\Delta\omega) \sigma_{x,y} U^\dag(\epsilon,\Delta\omega)\right\}
\end{equation}

\begin{equation}
  O_{x} = 1 - 2\cos^{2}(\Delta\omega\tau) \epsilon^{2} + \mathcal{O}(\epsilon^{4})
\end{equation}

\begin{equation}
  O_{y} = 1 - \frac{1}{8} \sin^{2}(2\Delta\omega\tau) \epsilon^{4} + \mathcal{O}(\epsilon^{5})
\end{equation}
Notice that if the rotation axis and the initial magnetization are aligned then the pulse error appears only in the $4^{\mathrm{th}}$ order of $\epsilon$, while for the original CP sequence the pulse error appears already in $2^{\mathrm{nd}}$ order.

The value of the CPMG sequence is that it simplifies the spin dynamics when applied over a distribution of Hamiltonians. For most applications, the relevant distribution is a spatially dependent spread in off-resonance frequencies, $\Delta\omega$, and scaled RF field amplitudes, $\omega_1$, relative to a nominal value ($\omega_1 = 1$). The overall dynamics then correspond to a convex operator sum over the classical probability distribution, $P(\Delta\omega,\omega_{1})$:

\begin{equation}\begin{split}
  \rho^{(n)}_{\mathrm{out}} = \int & P(\Delta\omega,\omega_{1}) \left[U_{\mathrm{cycle}}(\Delta\omega,\omega_{1})\right]^{n} \\& \times \rho_{\mathrm{in}} \left[U^\dag_{\mathrm{cycle}}(\Delta\omega,\omega_{1})\right]^{n} \, \mathrm{d}\Delta\omega \, \mathrm{d}\omega_{1},
\end{split}\end{equation}
where $\rho^{(n)}_{\mathrm{out}}$ is the density operator after the application of $n$ cycles (2$n$ pulses). The averaging must be undertaken after the propagator for each element of the distribution is raised to the $n^{\mathrm{th}}$ power. An important result in the design of CPMG sequences is that the projection of the rotation axis of not only the refocusing pulse but also the cycle propagator, $U_{\mathrm{cycle}}$, onto the initial magnetization must be as large as possible \cite{Hurlimann:01a}. If this is true for each element of the distribution, it will be true when the result is averaged over $P(\Delta\omega,\omega_1)$. Additionally, it will be true for all $n$, as the rotation axis will remain unchanged when the individual propagators are raised to the $n^{\mathrm{th}}$ power - corresponding to repeated applications of the cycle. This requires the cycle propagator for each element of $P(\Delta\omega,\omega_{1})$ to be expressable as 

\begin{equation}
  U_{\mathrm{cycle}}(\Delta\omega,\omega_{1}) = e^{-i\frac{\theta(\Delta\omega,\omega_{1})}{2}\sigma_{y}},
\end{equation}
such that all spins undergo a y-axis rotation of any angle over the cycle. This requirement places demands on the effective rotation axis and nutation angle of the refocusing pulses in the CPMG cycle.

The inherent robustness of the CPMG sequence to pulse nutation errors is reflected in the rotation axis of the cycle propagator. As stated previously, when the pulse rotation axis is taken to be aligned with the initial magnetization, the cycle rotation axis error appears to 4$^{\mathrm{th}}$ order in small deviations from a $\pi$ nutation angle. However, when the nutation angle is taken to be exactly $\pi$ radians, the cycle rotation axis error appears as the cosine of small deviations from a y-axis rotation, first appearing to 2$^{\mathrm{nd}}$ order. It is apparent that a deviation of the refocusing pulse from a perfect y-axis rotation is more detrimental to a successful CPMG sequence than is an identical deviation from a perfect $\pi$ nutation angle. 

There is some flexibility in the precision of the rotation characterisitics of the pulse, if a certain amount of signal loss is tolerable. The signal for a particular value of $\Delta\omega$ and $\omega_1$ after $k$ echoes is given by:

\begin{equation}
  M_y = (-1)^k \cos(k\delta)(1-r_y^2)+r_y^2
\end{equation}
where $\delta$ is the deviation of the half-cycle nutation angle from $\pi$ radians and $r_y$ is the y-component of the half-cycle rotation axis for the isochromat being considered. In order to retain 99$\%$ of the initial magnetization, for example, we require $r_y$ to be 0.995 for each isochromat, implying the pulse rotation axis must be within roughly 6 degrees of the initial magnetization, while the pulse nutation angle only need be within roughly 12 degrees of 180 degrees. However, since it is desirable to maximize the retained signal, it is necessary to require that the refocusing pulses be as close to a $\pi$ rotation about the y-axis as possible, leading to $\theta(\Delta\omega,\omega_1) \approx 2\pi$ and $U_{\mathrm{cycle}}(\Delta\omega,\omega_1) \approx $ I (identity) for all $\Delta\omega$ and $\omega_1$. For input states orthogonal to $\sigma_y$, the deviation from identity is cumulative for repeated cycle applications and leads to pulse-error-induced dephasing of the input state.

\section{Methods}
\label{sec:Methods}

\subsection{Optimal Control Theory for Unitary Pulse Design}
\label{subsec:octinnmr}

It is well-known that better compensation for static and RF field inhomogeneity may be achieved by using composite pulses \cite{Levitt:86a}, adiabatic pulses \cite{BaumPines:85a,KupceFreeman:95a,GraafGarwood:95a,GarwoodDelaBarre:01a}, and shaped pulses \cite{Freeman:98a}. These pulses achieve superior performance by increasing the number of degrees of freedom of the pulse shape. As noted previously \cite{SmithShaka:01a}, the most general pulse possible is simply a list of amplitudes and phases that do not necessarily adhere to a simple functional form. However, optimizing a general waveform can be challenging, especially for long waveforms with a corresponding large parameter space. A simulated annealing algorithm has been successfully applied to the design of pulses containing many periods of amplitude and phase variation \cite{Freeman:98a,GeenFreeman:91a}. Convergence was achieved by reducing the number of degrees of freedom of the pulse shape by representing the waveform as a sum of Fourier components. OCT techniques based on gradient search algorithms have proven exceptionally useful to efficiently find solutions in a large parameter space \cite{ConnollyMacovski:86a,MaoAndrew:86a,RosenfeldZur:96a,SkinnerGlaser:03a}. OCT is a well-established method to determine locally optimal solutions in a multivariate space, subject to a cost-functional \cite{Pontryagin:62a,Sage:68a,Bryson:75a,Pinch:93a}. Although these solutions correspond to local optima, these pulses have been demonstrated to yield excellent results and have found application in both NMR \cite{SkinnerGlaser:06a,SkinnerGlaser:04a} and ESR \cite{HodgesYang:08a}.

A great deal of effort has been devoted in the past to using OCT techniques to optimize excitation and inversion pulses which achieve a single state-to-state transformation with a high degree of accuracy \cite{ConnollyMacovski:86a,MaoAndrew:86a,SkinnerGlaser:03a,KobzarLuy:04a}. The CPMG sequence, however, requires pulses that perform a universal rotation, acting as a single unitary operation on all input states. It has been shown previously that state-to-state pulses can be made into universal rotation pulses, yielding refocusing pulses of twice the duration \cite{LuyGlaser:05a}. For example, given a 90$^\circ$ excitation pulse, a universal rotation pulse may be constructed by first applying a phase-reversed version of the excitation pulse, followed immediately by a time reversed version of the excitation pulse. While this symmetrization procedure leads to good refocusing pulses suitable for application in the CPMG sequence, we focus here on searching for general, non-symmetrized pulses. Expanding the set of potential solutions could possibly lead to pulses with better performance.

A treatment of the problem for a single spin species in SU(2) - neglecting relaxation and diffusion effects - involves numerically inverting the Liouville-von Neumann equation, which describes the evolution of a density operator, $\rho$, under the action of a generally time-dependent Hamiltonian:

\begin{equation}
	\frac{\mathrm{d}\rho}{\mathrm{d}t} = -i\left[\rho,H(t)\right].
\end{equation}
The general solution to this equation may be written in terms of a unitary propagator, $U(t)$: 

\begin{equation}
  \rho(t) = U(t) \, \rho_{0} \, U^\dag(t)
\end{equation}

\begin{equation}
    U(t) = \mathrm{T} \, e^{-i \int H(t) \, \mathrm{d}t},
    \label{eq:propagator}
\end{equation}
where T is the Dyson time ordering operator. The challenges in pulse design stem from the difficulty of finding a time-dependent Hamiltonian which implements a useful approximation to a desired unitary operation. The dominant Hamiltonian generating the dynamics for an ensemble of isolated spin-$\frac{1}{2}$ nuclei consists of the Zeeman interaction with the applied static, $B_0$, field and the resonant interaction with the applied RF, $B_1$, field:

\begin{equation}\begin{split}
	H(t) = & \frac{\Delta\omega}{2} \sigma_{z} + A(t)\frac{\omega_{1}}{4} \\& \times e^{-i \left(\omega_t t + \phi(t)\right) \sigma_{z}/2} \sigma_{x} e^{i \left(\omega_t t + \phi(t)\right) \sigma_{z}/2}.
	\label{eq:hamiltonian}
\end{split}\end{equation}                   
Here the OCT pulse is a time-dependent modulation of the RF amplitude, $A(t)$, and phase, $\phi(t)$, applied at a transmitter frequency $\omega_t$. In our notation, $\omega_1$ is a dimensionless scaling factor of the RF amplitude. Our goal is to find time sequences of the control parameters, $\{A(t),\phi(t)\}$ which, taken over the set of Hamiltonians determined by $P(\Delta\omega,\omega_{1})$, correspond to action which is sufficiently 'close' to a desired transformation, as measured by an appropriate performance functional.

The choice of performance functional is critical in the design process. As we are optimizing over a classical probability distribution, an appropriate performance functional is given by the convex weighted sum of the unitary fidelities for each member of the distribution: 

\begin{equation}
  \widetilde{\Phi} = \sum_{\Delta\omega,\omega_1}{P(\Delta\omega,\omega_{1}) \, \Phi_{\Delta\omega,\omega_{1}}(U_{\mathrm{pulse}},U_{\mathrm{targ}})},
  \label{eq:avgfidel}
\end{equation}
where
\begin{equation}\begin{split}
  \Phi_{\Delta\omega,\omega_{1}} & (U_{\mathrm{pulse}},U_{\mathrm{targ}}) = \\& \frac{\left| \mathrm{Tr} \left\{ U_{\mathrm{pulse}}(\Delta\omega,\omega_{1}) \, U_{\mathrm{targ}}^\dag \right\} \right|^{2}}{4}.
  \label{eq:unitaryfidel}
\end{split}\end{equation}
The performance functional (\ref{eq:unitaryfidel}) is a standard metric for the distance between two unitary operators in SU(2) and is equivalent to the average of the correlations of a complete set of input states evolved under $U_{\mathrm{pulse}}$ and $U_{\mathrm{targ}}$, respectively \cite{FortunatoCory:02a}. 

The behavior of (\ref{eq:unitaryfidel}) with respect to $\Delta\omega$ and $\omega_1$ is sufficient to ensure the satisfaction of the CPMG criteria. Consider a general rotation in SU(2),

\begin{equation}
  U(\theta,\widehat{r}) = e^{-i\frac{\theta}{2} \widehat{r} \cdot \vec{\sigma}},
\end{equation}
with nutation angle $\theta$ about an axis given by the unit vector $\widehat{r}$. The corresponding fidelity to a $\pi$ rotation about the y-axis is

\begin{equation}
  \Phi(U,e^{-i \frac{\pi}{2} \sigma_{y}}) = \sin^{2}{\frac{\theta}{2}} r^{2}_{y} = \sin^{2}{\frac{\theta}{2}} \cos^{2}{\phi},
  \label{eq:fidelroterr}
\end{equation}
where $\theta$ is, again, the nutation angle of the pulse and $\phi$ is the angle between $\widehat{r}$ and $\widehat{y}$. Small deviations from a $\pi$ rotation about the y-axis cause the fidelity to decrease quickly, ensuring that high-fidelity correponds to the satisfaction of CPMG criteria (i) and (ii). Additionally, as noted in CPMG criteria (iii), the requirement that it is more important to minimize $\phi$ than it is to minimize $\theta$ is reflected in the behavior of the fidelity. It is useful to note that, in our approach, it is easy to substitute a particular functional for a different one. While our chosen performance functional exhibits the general behavior we desire, we cannot rule out the existence of another functional which may be more sensitive to deviations in $\phi$ while allowing more flexibility in $\theta$. Such a functional could possibly lead to more accurate solutions with an enhanced bandwidth. However, this aspect of OCT's versatility has not been explored in this work.

\subsection{The GRAPE Algorithm}    

In order to find locally optimal pulses which achieve high fidelity over $P(\Delta\omega,\omega_1)$ we utilize the efficient optimal control GRadient Ascent Pulse Engineering (GRAPE) algorithm developed by Khaneja and coworkers. Here we provide only the details of the algorithm relevant to this work. Complete details and background of the algorithm may be found in the references \cite{KhanejaGlaser:05a,GlaserGriesinger:98a}.

We consider here OCT pulses that are defined to be piecewise constant over a number of intervals, $N$, each of length $\Delta t$, which yields an overall pulse length of $T = N \Delta t$. During each interval, the amplitude, $A$, and phase, $\phi$, of the RF modulation is set to constant values, such that the pulse is described as a list $\left\{ A_{j},\phi_{j} \right\}$, where $j$ runs from 1 to $N$. The GRAPE algorithm proceeds by choosing an initial collection of RF amplitudes and phases, calculating the Hamiltonian, propagator, and fidelity (eqs. \ref{eq:hamiltonian}, \ref{eq:propagator}, and \ref{eq:unitaryfidel}) for each element of $P(\Delta\omega, \omega_1)$, then updating the controls based on the respective gradients of the fidelity. This process repeats until either a desired value of $\widetilde{\Phi}$ is reached or the improvement in fidelity from one iteration to the next is less than a preset threshold value, indicating a local optimum has been reached. As we are optimizing over a classical disribution, the fidelities and gradients were calculated for each isochromat, then averaged based on the weightings defined by the distribution. In the notation of \cite{KhanejaGlaser:05a} the fidelity and corresponding gradients to first order in $\Delta t$ for each isochromat are

\begin{equation}
  \Phi = \left|\left\langle U_{\mathrm{targ}} | U(T)\right\rangle\right|^{2} = \left\langle P_{j} | X_{j} \right\rangle\left\langle X_{j} | P_{j} \right\rangle
\end{equation}

\begin{equation}
  \frac{\delta\Phi}{\delta u_{k}(j)} = -\mathrm{Re}\left\{\left\langle P_{j} | i\Delta\tau H_{k} X_{j} \right\rangle\left\langle X_{j} P_{j} \right\rangle \right\}
  \label{eq:gradients}
\end{equation}
where $P_{j} = U_{j+1}^{\dag} \ldots U_{N}^{\dag} U_{\mathrm{targ}}$ and $X_{j} = U_{j} \ldots U_{1}$ are the backward and forward representations of the total pulse propagator at the $j_{th}$ time step, and $u_{1}(j) = A_{j}\cos\left(\phi_{j}\right)$, $u_{2}(j) = A_{j}\sin\left(\phi_{j}\right)$, $H_{1} = \sigma_{x}$, and $H_{2} = \sigma_{y}$ are the values and representation of the RF control parameters in Cartesian coordinates. The timestep, $\Delta t$ must be chosen much smaller than the inverse magnitude of the Hamiltonians in order for the truncation of the expansion of the gradients (eq. \ref{eq:gradients}) to be valid.

The updating step for RF amplitudes may be written as

\begin{equation}
   u_{k}^{n+1}(j) = u_{k}^{n}(j) + \epsilon \frac{\delta\widetilde{\Phi}}{\delta u_{k}^{n}(j)}
\end{equation}
where $\epsilon$ is an adjustable step-size and $n$ refers to the iteration number of the algorithm. The step-size is adjusted using a line-search at the end of each iteration in order to maximize the increase in $\widetilde{\Phi}$ without making large deviations in the control parameters between iterations, hindering convergence. Also, in order to prevent the RF amplitude from exceeding a predefined value, any pulsing periods which exceed the maximum are simply reset to the maximum at the end of each iteration. 

\subsection{Iterative Optimization Strategy}
In the absence of constraints, the GRAPE algorithm guarantees that our optimized pulse parameters will deterministically converge to the nearest local maximum of average fidelity. The value of the average fidelity at this local maximum depends on the structure of the control landscape - a geometric representation of the value of the performance functional as a function of the control parameters. By definition, the control landscape for our chosen performance functional is given by the weighted convex sum of the control landscapes for each member of the distribution, $P(\Delta\omega, \omega_1)$. The stucture and dimensionality of the control landscape is determined by our choice of $T$, $\Delta t$, and $P(\Delta\omega, \omega_1)$. Limiting the maximum instantaneous RF power constrains the ability of the algorithm to reach a local optimum from a given random initial guess of the pulse parameters, $\left\{ A_{j},\phi_{j} \right\}_{\mathrm{init}}$  

In idealized situations, two key results of quantum control theory provide some insight into the structure of the control landscape, and thus, the level of control we can expect to achieve. In general, though, very little may be concretely said about the structure of the control landscape in quantum optimizations, which makes it extremely difficult to find the best possible (global) solution. It has been shown that in the absence of decoherence every local optima is exact \cite{RabitzRosenthal:04a,ChakrabartiRabitz:07a}, and even in the presence of a static, classical distribution of Hamiltonians there exist solutions which provide exact implementations of the desired unitary operation for every member of the ensemble (see online material for \cite{RabitzRosenthal:04a}). However, these results only apply to situations where there are no constraints on pulse length and amplitude and the pulse shape is taken to be changing continuously as a function of time. Imposing these constraints, as we do in this work, implies that our solutions will not reach the desired optimum over the distribution, and that no optimum will correspond to an exact implementation of the desired unitary operator. When optimizing a particular initial guess to the nearest optimum point, then, we are sampling from the set of non-degenerate local optima without any guarantee of the performance being close to the global maximum.

We tried two systematic methods to investigate the maximum achievable pulse performance for our techniques and constraints, each of which produced similar results but differed in efficiency and information gained about the structure of the control landscape. The first method involved choosing a particular width of the distribution and optimizing hundreds of random initial guesses to obtain a histogram of achievable fidelities. The best of these was then taken as the maximum performance we could achieve for the distribution in question. The width of the distribution was then increased, and the process repeated, until the highest achievable fidelity was no longer satisfactory. Our results for a single width of the distribution confirmed the behavior seen in \cite{KobzarLuy:04a}. A wide range of optimized fidelities were obtained, with the majority clustered around higher values. The results of these optimizations tell us that the static collective landscape for our problem contains many varieties of local optima that may trap gradient algorithms, but the performance of most of these optima is similar and normally correspond to good control.

To gain further insight into the way the collective landscape changes as new elements are added to the distribution, we also implemented a new method, where a single initial guess is systematically optimized over distributions of increasing width. We first attempted to maximize the bandwidth of the pulse with respect to static field inhomogeneity, while neglecting RF-inhomogeneity (i.e. setting $\omega_1 = 1$). The target bandwidth of the pulse was defined to be $2\Delta\omega_{\mathrm{max}}$, where the uniform distribution was defined as $\overline{P}(\Delta\omega) = 1/M$ for $M$ values of $\Delta\omega = (-\Delta\omega_{\mathrm{max}}, ..., -2\delta\omega, -\delta\omega, 0, \delta\omega, 2\delta\omega, ..., \Delta\omega_{\mathrm{max}})$. The value of $\delta\omega$ is determined by the length of the pulse in order to ensure that if high fidelity is achieved at each point in the distribution, the fidelity taken continuously in between will remain high. We found that a value of $\frac{1}{4T}$, where T is the length of the pulse, is sufficient. 

The iterative optimization procedure started with an on-resonance optimization, $\overline{P}_1(\Delta\omega = 0) = 1$, where the results of quantum control theory dictate that unit fidelity is always achievable regardless of initial guess. The resulting optimized pulse was then used as the initial guess for an optimization over 3 isochromats, $\overline{P}_2(\Delta\omega = -\delta\omega$, 0, $\delta\omega) = 1/3$, and allowed to run until a local maximum of the new collective landscape was found. A small amount of randomization was added such that the the distance between isochromats was not exactly equal. After a pulse was optimized over 3 isochromats, that pulse was used as an initial guess for 5 isochromats, $\Delta\omega = (-2\delta\omega$, $-\delta\omega$, 0, $\delta\omega$, $2\delta\omega)$, and so on. 

The iterative process was terminated when the average fidelity achieved over the distribution dropped below roughly 0.9. We found that by using this process, regardless of our initial guess, a series of convergent pulses with an associated fidelity vs. static inhomogeneity bandwidth curve was generated (Fig. \ref{fig:IterativeOpt}) and a satisfactory result was obtained in only the order of tens of optimizations. The characteristics of this curve varied depending on the initial guess, but always yielded usable solutions. In order to account for RF-inhomogeneity a pulse from the related set generated by the iterative optimization was chosen to be reoptimized. Without changing the bandwidth of the pulse with respect to static field inhomogeneity, a certain amount of RF-inhomogeneity was added. For example, if we chose a pulse that was optimized over 41 values of $\Delta\omega$ and we added in $\pm 10 \%$ RFI given by $\omega_{1} = 0.9, 0.95, 1, 1.05, 1.1$, the resulting distribution $\overline{P}(\Delta\omega,\omega_{1}) = 1/204$ would have 41*5 = 204 elements.

The success of this particular approach suggests that the structure of the collective landscape characterizing the average fidelity does not change abruptly as the width of the distribution is increased slowly. A good solution for a given distribution width should therefore remain in the immediate neighborhood of a local optimum with high fidelity for a slightly wider distribution width. However, concrete statements about the structure of the control landscape are the subject of quantum control theory, and are outside the scope of this present work. Additionally, while we are not aware of any published bound in the NMR literature on the globally maximal pulse performance subject to the constraints considered here, we note that the current problem is related to problems in holonomic control that are naturally posed in SO(3). As such, a published bound may exist in the literature on holonomic control.

\section{Results}
\label{sec:Results}
We present specific results obtained by optimizing the fidelity of OCT refocusing pulses constrained to be 1 ms duration and with a maximum instantaneous RF amplitude of $A_{\mathrm{max}} / 2\pi = 5$ kHz. Our pulses were defined piecewise constant over 100 intervals of 10 $\mu$s duration and the 200 degrees of freedom in the problem - 100 periods of varying pulse amplitude and phase - were optimized iteratively over a uniform distribution using the GRAPE algorithm at each iteration. A 6 $\mu$s delay before and after the pulse was incorporated into the optimization to account for hardware switching times in back-to-back applications of our pulses. These delays are not directly part of the pulse waveform, but must be included in simulations and experiments in order for the pulse to function properly.

\begin{figure}[!hbt]
  \begin{center}
    \includegraphics[scale=1,keepaspectratio]{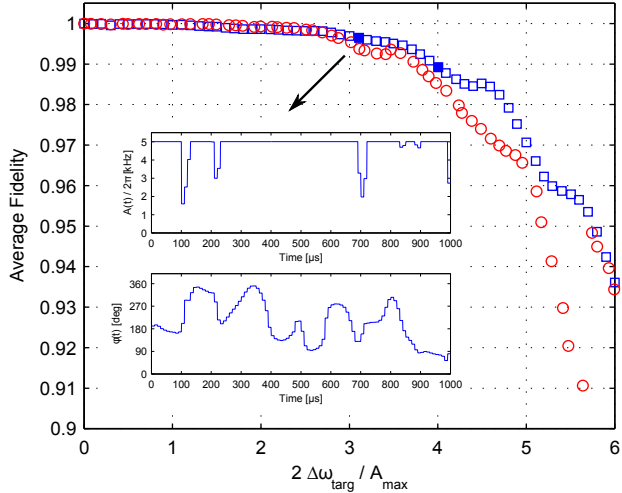}
  \end{center}
    \caption{{\small [color online and in print] Two representative examples of the iterative optimization procedure utilized in this study. The results shown by the red circles and blue squares correspond to two iterative series of pulses derived from two different random initial guesses, respectively. The pulses are iteratively optimized over an increasing target bandwidth of resonance offsets. We find that, regardless of the initial guess, the resulting curves of average fidelity versus target bandwidth are similar, indicating that the pulses for the different realizations have similar maximum bandwidth. The filled in squares represent the pulses mentioned in the text, with the inset showing the temporal profile of the pulse chosen for extended analysis. High-resolution pulse profile and parameter list is available in supplementary material.}}
  \label{fig:IterativeOpt}
\end{figure}

\begin{figure}[!hbt]
  \begin{center}
    \includegraphics[scale=1,keepaspectratio]{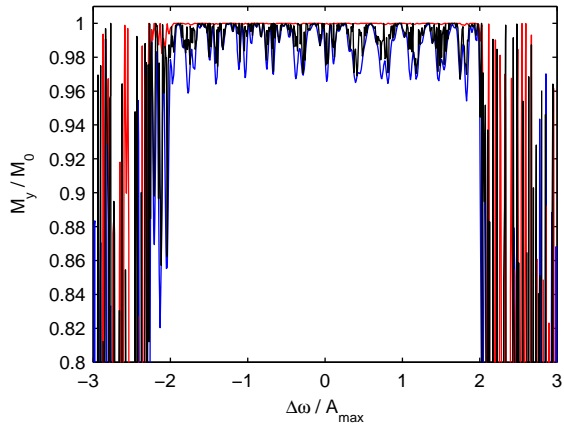}
  \end{center}
    \caption{{\small [color online and in print] Calculated magnetization as a function of resonance offset for the first (blue), second (red), and 500th (black) echoes of a CPMG sequence using the 1 ms OCT refocusing pulse optimized over $\left| \Delta\omega \right| \leq $ 2 $A_{\mathrm{max}}$. The simulation assumes a perfect excitation pulse, no relaxation, and uniform RF. The time between refocusing pulse applications is $2\tau = 2$ ms. The magnetization is very nearly retained over the entire optimization range, and shows no additional degradation for higher echo numbers.}}
    \label{fig:echovis}
\end{figure}

Using the iterative optimization procedure we generated a series of related pulses, terminating the series when the average fidelity dropped below roughly 0.9 for the target bandwidth being optimized. Figure \ref{fig:IterativeOpt} displays two representative curves of the average fidelity of the optimized pulses as a function of target bandwidth ($\left| \Delta\omega \right| \leq \Delta\omega_{\mathrm{targ}}$). Not surprisingly, for small target bandwidths it is possible to find pulses with an average fidelity very nearly unity; unit fidelity was achieved only for the on-resonance optimization. As the target bandwidth is increased, the average fidelity begins to drop, with the detailed characteristics of the curve dependent upon the initial guess. We found that the target bandwidth at which the average fidelity drops below 0.99 was nearly the same for all initial guesses we tried. We took this point to represent the maximum bandwidth achievable before pulse performance is significantly affected.             
Based on this performance criterium, the best refocusing pulse we could find in the absence of RF inhomogeneity (RFI) is marked in fig. \ref{fig:IterativeOpt} and has an average fidelity $\widetilde{\Phi} = 0.989$ over a total bandwidth of 4 $A_{\mathrm{max}}$. Figure \ref{fig:echovis} shows the calculated response of a CPMG sequence using this refocusing pulse, in the absence of relaxation and RFI. The response is almost entirely maximal across the entire optimization range with refocusing of over 96$\%$ of the magnetization at any offset. This performance is retained at higher echo numbers. This confirms that optimizing the average fidelity maximizes the echo visibility and allows the generation of a large number of echoes.

When RFI is included in the optimizations, a trade-off between achievable resonance offset bandwidth and RFI compensation becomes apparent. For modest RFI of $\pm 10\%$ ($\omega_1 = 0.9 - 1.1$), the average pulse fidelity of the pulse optimized over a static offset bandwidth of 4 $A_{\mathrm{max}}$ drops to 0.972, while a pulse optimized over a static offset bandwidth of 3.2 $A_{\mathrm{max}}$ has average fidelity of 0.982. In the absence of RFI, the latter pulse has average fidelity of 0.996 over the respective bandwidth. We chose this pulse for further analysis. It's temporal profile is shown in the inset of fig. \ref{fig:IterativeOpt}. The RF tends to remain on for nearly the entire pulsing time while the phase is non-trivially varied in a non-symmetric way.

\begin{figure*}[!bt]
  \begin{center}
    \subfigure[Fidelity]{\includegraphics[scale=1,keepaspectratio]{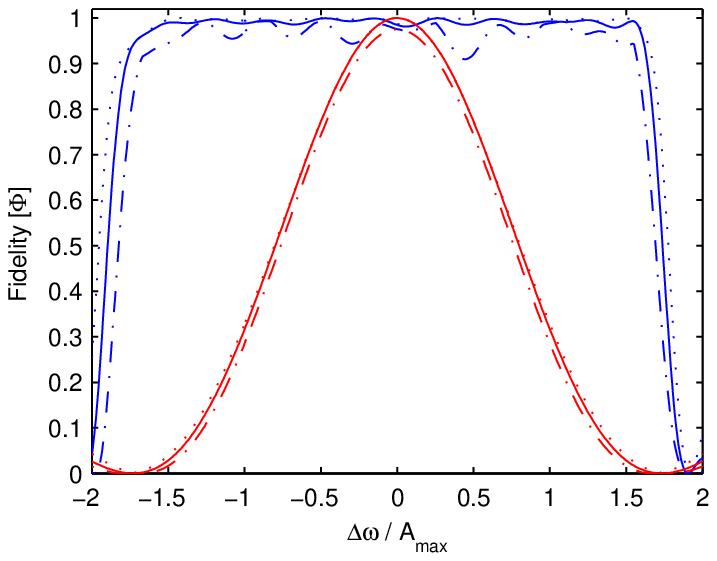}}
    \subfigure[Angle From XY-Plane]{\includegraphics[scale=1,keepaspectratio]{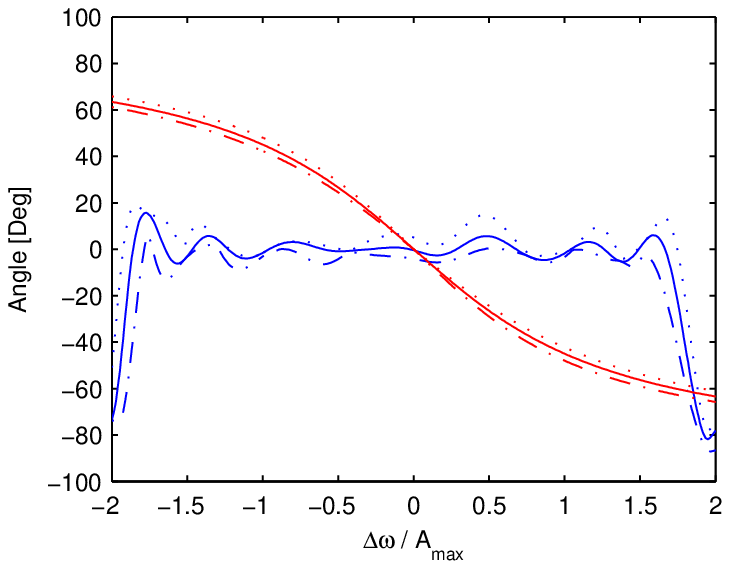}}
    \subfigure[Angle From Y-axis]{\includegraphics[scale=1,keepaspectratio]{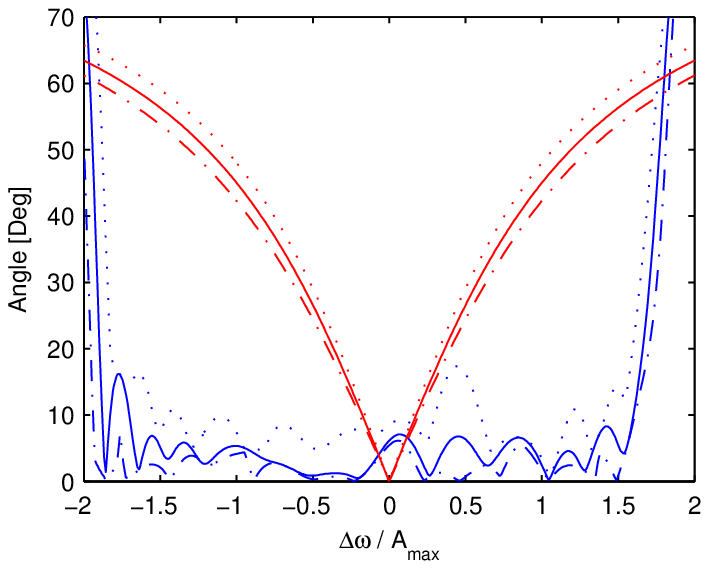}}
    \subfigure[Nutation Angle]{\includegraphics[scale=1,keepaspectratio]{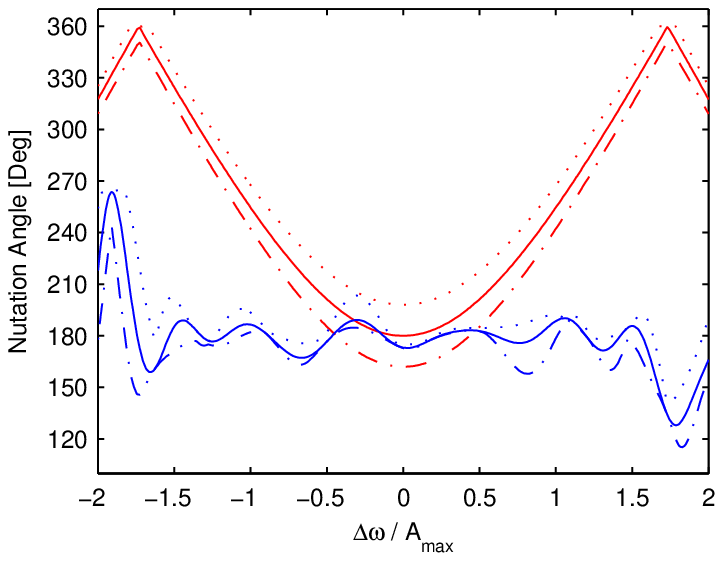}}
  \end{center}
    \caption{{\small [color online and in print] Fidelity and CPMG criteria as a function of resonance offset for the 1 ms OCT refocusing pulse (Blue) optimized over $\left| \Delta\omega \right| \leq$ 1.6 $A_{\mathrm{max}}$ and $\omega_1 = 0.9 - 1.1$, as compared to a 100 $\mu$s hard pulse (Red). The solid lines indicate the response for uniform RF ($\omega_1 = 1$), while the dotted (dash-dotted) lines indicate the maximum (minimum) angle over the range of $\omega_1 = 0.9 - 1.1$. The rotation axis for the OCT pulse stays within 15$^\circ$ of the initial magnetization over the optimized distribution and the nutation angle remains within 30$^\circ$ of 180$^\circ$.}}
    \label{fig:cpmgsatisfy}
\end{figure*}

To verify that high average fidelity leads to the desired pulse performance, we must consider the fidelity and CPMG criteria of our OCT refocusing pulse as a function of resonance offset. Given the relation between fidelity and pulse rotation errors (eq. \ref{eq:fidelroterr}), an average fidelity close to unity implies that for every value within the considered $B_0-B_1$ distribution, the net action of the optimized pulse is very close to a $\pi$ rotation around the y-axis. Figure \ref{fig:cpmgsatisfy} demonstrates that this is indeed the case. For comparison, we include the corresponding quantities for a standard hard $\pi$ pulse of the same maximum RF amplitude (duration 100 $\mu$s). Note that, in general, the deviation of the rotation for our OCT pulse from a y-axis rotation is smaller than the deviation from a $\pi$ nutation angle, as desired.

\subsection{Experimental Verification}
\label{subsec:experimental}
In order to verify that our OCT pulses perform as expected in experiment, we performed CPMG measurements on a sample of 90\% D$_{2}$O / 10\% H$_{2}$O. Roughly 3 mM copper-sulfate (CuSO$_{4}$) was added to the sample in order to obtain a relaxation time of T$_{2} \approx$ 270 ms at 300 MHz proton resonance. RF inhomogeneity was measured to extend out to $\pm$20$\%$, with the majority of the field strengths concentrated in the $\pm$10$\%$ range. We performed CPMG measurements using the OCT refocusing pulse shown in the inset of fig. \ref{fig:IterativeOpt}, having a total duration of 1 ms and a maximum RF amplitude $A_{\mathrm{max}}/2\pi$ = 5 kHz, with a time between refocusing pulses of 2$\tau$ = 20 ms. The 90$^\circ$ excitation pulse was an on-resonance rectangular hard pulse of amplitude 31.25 kHz to ensure the sequence performance was limited only by the refocusing pulses. To test the performance under off-resonance conditions, we systematically varied the offset of the proton transmitter frequency from the Larmor frequency in 100 Hz increments in the range from -10 kHz to 10 kHz (sequences run in an applied gradient field are included in the supplementary material for verification). Figure \ref{fig:expcpmg}a shows the amplitudes of the first 13 echoes of a CPMG sequence with OCT refocusing pulses as a function of offset frequency. As a comparison, we show in fig. \ref{fig:expcpmg}b the first 10 echoes of a CPMG sequence using standard, hard refocusing pulses of 5 kHz amplitude. The measurements confirm that, for the OCT refocusing pulses, pulse errors do not contribute significantly to the observed decoherence and the echo visibility is maximal over the entire optimized range of $\left| \Delta\omega \right| \leq 1.6$ $A_{\mathrm{max}}$. 

\begin{figure}[!hbt]
  \begin{center}
    \subfigure[OCT Pulses]{\includegraphics[scale=1,keepaspectratio]{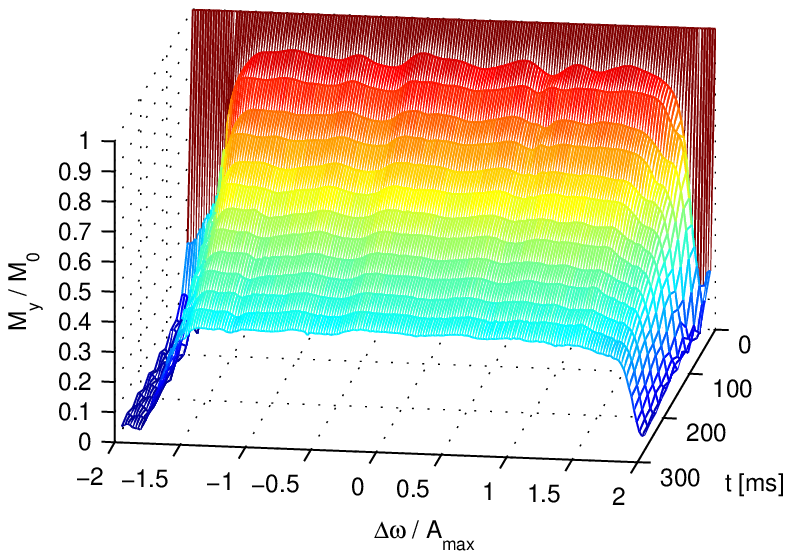}}
    \subfigure[Hard Pulses]{\includegraphics[scale=1,keepaspectratio]{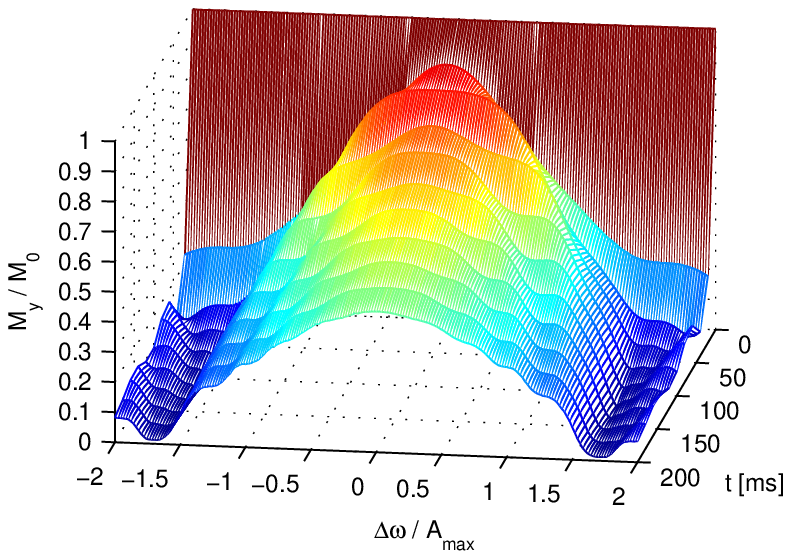}}
  \end{center}
  \caption{{\small [color online and in print] Experimental results of echo amplitudes generated by CPMG sequences with OCT refocusing pulses (a) and standard hard pulses (b) acquired at 300 MHz on a sample of CuSO$_4$ doped 90$\%$ D$_2$O / 10$\%$ H$_2$O with T$_2$  $\approx$ 270 ms. The results were acquired sequentially by changing the resonance offsets systematically over the range of $\left| \Delta\omega \right| \leq$ 2 $A_{\mathrm{max}}$. The time between refocusing pulses is $2\tau = 20$ ms. The CPMG sequence with OCT pulses generates a uniform response over resonance offsets in the range of $\pm$1.6 $A_{\mathrm{max}}$. The response of the excitation pulse is shown to be flat over the relevant range of resonance offsets.}}
  \label{fig:expcpmg}
\end{figure}

\begin{figure}[!hbt]
  \begin{center}
    \subfigure[$\Delta\omega / 2\pi$ = 100 Hz]{\includegraphics[scale=1,keepaspectratio]{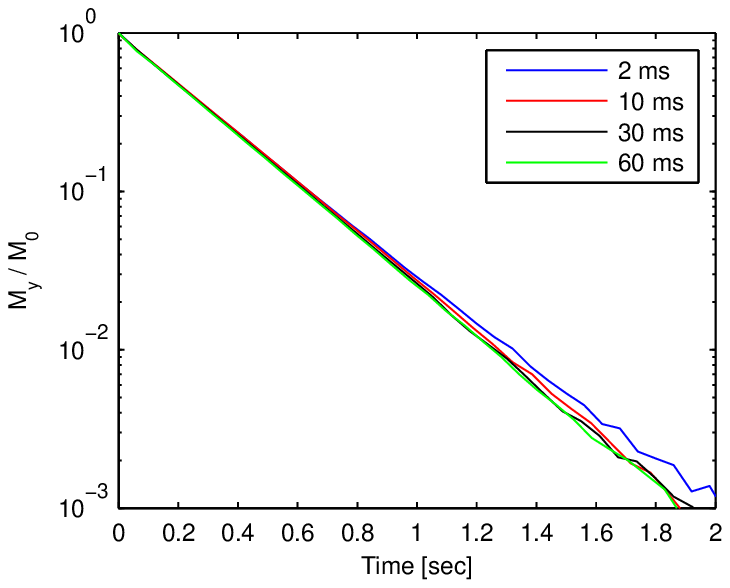}}
    \subfigure[$\Delta\omega / 2\pi$ = 7000 Hz]{\includegraphics[scale=1,keepaspectratio]{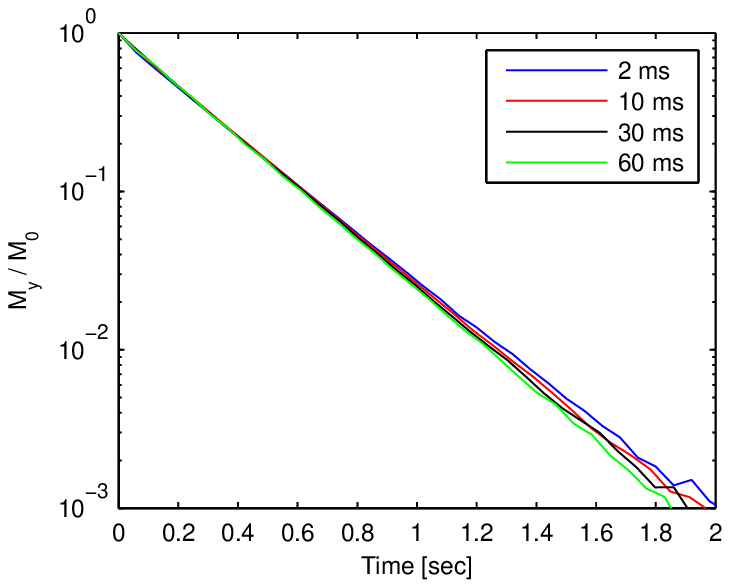}}
  \end{center}
    \caption{{\small [color online and in print] Comparison of measured echo amplitudes with different values of 2$\tau$ (time between refocusing pulses), as indicated in the legend, for a CPMG sequence with OCT refocusing pulses. The results show that a decrease in echo spacing does not lead to an increase in the relaxation rate, indicating that no additional relaxation is induced by the pulses. In fact, as the echo spacing is decreased and more pulses are applied, the measured relaxation time becomes longer. As discussed in the text, this is caused by T$_1$ effects during the application of the refocusing pulses. Additionally, the measured echo amplitudes show a small transient effect of roughly 1$\%$ amplitude, in agreement with expectation. }}
    \label{fig:expcompare}
\end{figure}

\begin{figure*}[!bt]
  \begin{center}
    \subfigure[$\Delta\omega / 2\pi$ = -5 kHz]{\includegraphics[scale=1,keepaspectratio]{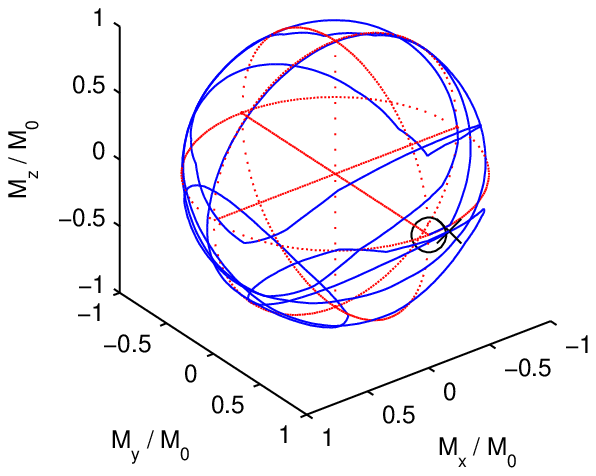}}
    \subfigure[$\Delta\omega / 2\pi$ = 5 kHz]{\includegraphics[scale=1,keepaspectratio]{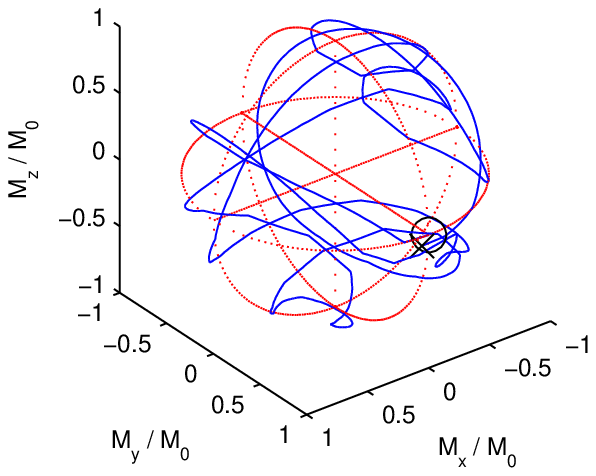}}
    \subfigure[$\Delta\omega / 2\pi$ = 0 ]{\includegraphics[scale=1,keepaspectratio]{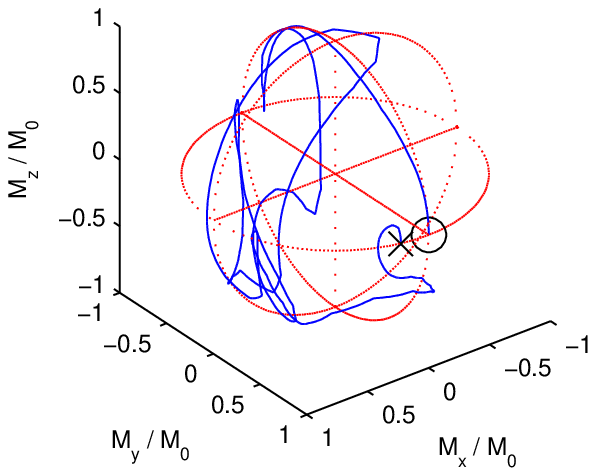}}
    \subfigure[$\Delta\omega / 2\pi$ = 1 kHz]{\includegraphics[scale=1,keepaspectratio]{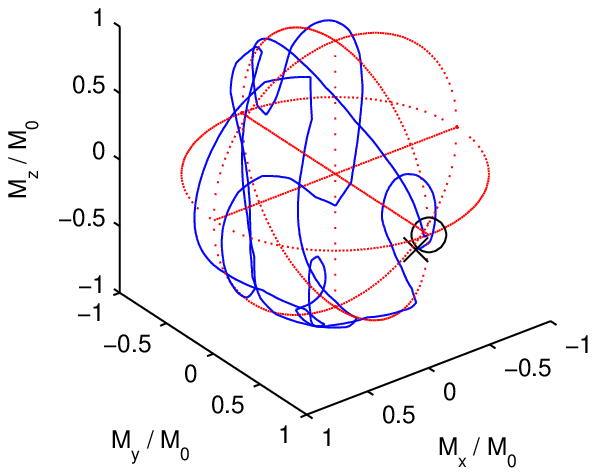}}
  \end{center}
    \caption{{\small [color online and in print] Bloch sphere trajectories during the OCT refocusing pulse for several values of resonance offset as applied to a $\sigma_{y}$ initial state. The trajectories for different isochromats vary significantly but result in nearly identical effective rotations. Roughly two-thirds of the magnetization is in the transverse plane during pulsing. The black 'o' denotes the initial state, while the black 'x' denotes the final state.}}
    \label{fig:blochtraj}
\end{figure*}

To further quantify the echo decay, we compared the amplitudes of the CPMG echoes as a function of time for different values of the time between OCT refocusing pulses, 2$\tau$, and for two particular values of resonance offset (fig. \ref{fig:expcompare}). The slope for each of these echo spacings is nearly identical, in agreement with our expectation. In addition, there is no evidence of additional pulse-induced relaxation decay when shorter echo spacings are used. In fact, our results suggest that when more pulses are applied in a given time period, T$_{\mathrm{2,eff}}$ becomes slightly longer. This is most likely due to T$_1$ being slightly longer than T$_2$. The complex trajectories taken on the Bloch sphere, as shown in fig. \ref{fig:blochtraj}, cause the magnetization to spend some amount of time away from the transverse plane. This implies that as more pulses are applied in a given time period, the effects of T$_1$ become increasingly important. While the exact proportion of time spent away from the transverse plane varies depending on resonance offset, in general the magnetization undergoes T$_1$ relaxation for roughly one third of the pulse duration. Using this approximation and a measured value of T$_2$ of roughly 270 ms (using a hard pulse CPMG sequence on-resonance), the measured T$_{\mathrm{2,eff}}$ of 282 ms for 2$\tau$ = 2 ms implies a T$_1$ of roughly 370 ms. These values predict an apparent T$_2$ for 2$\tau$ = 10 ms of 275 ms, consistent with our measurement. Similarly, the measured and expected values of T$_2$ for 2$\tau$ = 30 ms and 60 ms are close to 272 ms.  

\subsection{Comparison to Previous Pulses}
\label{subsec:comparison}
\begin{figure}[!hbt]
  \begin{center}
    \includegraphics[scale=1,keepaspectratio]{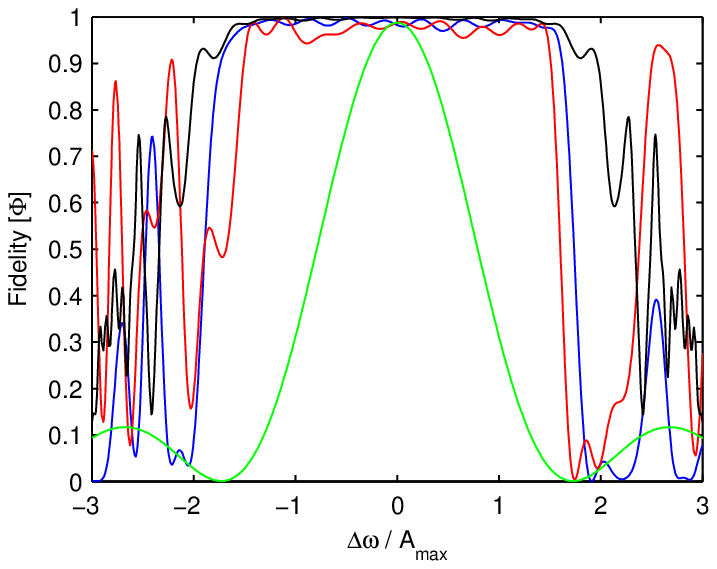}
  \end{center}
    \caption{{\small [color online and in print] Simulated fidelity as a function of resonance offset for our 1 ms OCT pulse optimized over $\pm$10$\%$ RFI and $\left|\Delta\omega\right| \leq$ 1.6 $A_{\mathrm{max}}$ (blue), a previously published OCT excitation pulse \cite{KobzarLuy:04a} optimized over $\pm$20$\%$ RFI and $\left|\Delta\omega\right| \leq$ 1.5 $A_{\mathrm{max}}$ and modified to be a 1.02 ms refocusing pulse (red), a 4 ms Chirp refocusing pulse (black) \cite{HwangGarwood:97a}, and a 100 $\mu$s hard pulse (green). All pulses have maximum RF amplitude $A_{\mathrm{max}}/2\pi$ = 5 kHz. The fidelities for each offset were averaged over $\pm$10$\%$ RFI ($\omega_1 = 0.9 - 1.1$). High-resolution time-domain profiles for each pulse are included in the supplementary material.}}
   \label{fig:pulcomparefidel}
\end{figure}

Figure \ref{fig:pulcomparefidel} compares the performance of our OCT pulse with previously published refocusing pulses by examining the fidelity as a function of resonance offset, averaged over $\pm$10$\%$ RFI. The Chirp pulse is a composite adiabatic pulse \cite{HwangGarwood:97a} composed of a base element Chirp adiabatic inversion pulse \cite{BohlenBodenhausen:89a,BohlenBodenhausen:93a,FuBodenhausen:95a} scaled to $A_{\mathrm{max}}/2\pi = 5$ kHz. The original pulse was 2 ms long, with a 60 kHz sweep width, 20$\%$ smoothing, and on-resonance adiabaticity of $Q_0 = 5$. Our scaled version changes the sweep width to 30 kHz and the pulse length to 4 ms, without changing $Q_0$. We found that, with the constraint of maximum RF amplitude, 4 ms was the minimum time required for the pulse to fulfill the adiabatic condition and function properly. Additionally, a previously derived 90$^\circ$ OCT excitation pulse was downloaded from the website mentioned in \cite{KobzarLuy:04a} and turned into a 1.02 ms refocusing pulse consisting of 1020 intervals of 1 $\mu$s duration by the method detailed in \cite{LuyGlaser:05a}. The excitation pulse was originally optimized over $\left| \Delta\omega \right| \leq 1.5$ $A_{\mathrm{max}}$ and $\omega_1 = 0.8 - 1.2$. The performance quality of the original excitation pulse with respect to $\Delta\omega$ and $\omega_1$ is retained by the symmetrized refocusing pulse.

As shown in the figure, each of the pulses considered performs very nearly as a $\pi)_y$ pulse over a range of resonance offsets of $\pm$1.6 $A_{\mathrm{max}}$. The Chirp pulse performs most closely to the desired behavior over the operating range, but at the expense of being four times longer than the OCT pulses. The performance of the pulse derived by symmetrizing a previously reported OCT excitation pulse is similar to the performance of our OCT pulse found by direct optimization without symmetrization. The fact that similar limits of performance for OCT pulses were obtained by different means and using many different initial guesses suggests that we may be near the global limit on bandwidth for the constraints considered, and that the performance of many solutions are clustered around the global maximum. However, a proper treatment of this claim requires further investigation and is outside the scope of this work.

While it is instructive to consider the performance criteria of a single refocusing pulse, it is difficult to use this information directly to infer the performance of the CP(MG) sequence over many pulses. To do so we must consider not only the errors which occur during a single pulse application, but how they correlate and evolve during the application of many pulses. In the next section, we take advantage of developments in the study of quantum channels to compactly describe the nature and severity of errors that occur during the CP(MG) sequence and quantify the effects those errors have on the performance of the sequence.

\section{Discussion and Analysis}
\label{sec:discussionanalysis}
To appreciate how the near satisfaction of the CPMG criteria for each isochromat leads to a successful CPMG sequence, we may consider the collective action averaged over $P(\Delta\omega,\omega_{1})$. As given by eq. 6, the overall dynamics correspond to a convex operator sum over the distribution. While the action is unitary for each isochromat, the overall action viewed as an effective map from $\rho_{\mathrm{in}}$ to $\rho_{\mathrm{out}}$ need no longer be unitary. In this case, representation via a superoperator is needed to provide a compact and complete description of the dynamics. 

\subsection{CPMG Superoperator}
\label{subsec:cpmgsup}
A superoperator is a general map that describes the dynamics of a physical system by defining how any valid input state is mapped to any valid output state \cite{Ernst:87a}. While a general quantum process may be conveniently described by a superoperator, ideally the map must be completely positive and trace-preserving. Requiring a map to be positive and trace-preserving ensures that valid quantum states (given by positive, Hermitean density operators) are mapped to other valid quantum states, and that probability is conserved. Demanding that the map be completely positive ensures that the same is true when the map is considered as a subsystem of a larger quantum system. When the superoperator is calculated over a distribution, as given in this case by field inhomogeneities, a completely positive map is only expected if the input state, $\rho_{\mathrm{in}}$, is uncorrelated over the distribution. It is important to note that in any experimental study of a physical system the superoperator will not be completely positive.  

The elements of the CPMG superoperator for $n$ cycle applications are determined by the action of the map on a spanning set of basis vectors in Liouville space. In the Pauli basis the superoperator for $n$ cycles of the CPMG sequence is 

\begin{equation}\begin{split}
  & \left\langle \widetilde{\sigma}_{\beta} \right| \widetilde{S}_{n} \left| \widetilde{\sigma}_{\alpha} \right\rangle = \frac{1}{2} \mathrm{Tr} \bigg{\{} \sigma_{\beta} \int P(\Delta\omega,\omega_{1}) \\& \times \left[ U_{\mathrm{cycle}}(\Delta\omega,\omega_{1}) \right]^{n} \sigma_{\alpha} \left[ U_{\mathrm{cycle}}^\dag(\Delta\omega,\omega_{1}) \right]^{n} \\& \, \mathrm{d}\Delta\omega \, \mathrm{d}\omega_{1} \bigg\}. 
  \label{eq:pobasissup}
\end{split}\end{equation}
Here $\sigma_{\alpha,\beta} = \left\{I,\sigma_{x},\sigma_{y},\sigma_{z} \right\}$ are the usual spin-$\frac{1}{2}$ Pauli operators and $\left| \widetilde{\sigma}_{\alpha,\beta} \right\rangle$ are the columnized version of the operators obtained by stacking the first column of the respective Pauli matrix on top of its second column \cite{Havel:03a}. The columnized Pauli operators are basis vectors in Liouville space. 

While the unitary dynamics for a single spin-$\frac{1}{2}$ are describable in a 2-dimensional Hilbert space, the superoperator is described in a 4-dimensional Liouville space. Notice that, because the superoperator is averaged over a physical distribution, we must describe a superoperator individually for each $n$. This is equivalent to the statement made in section \ref{sec:cpmgcriteria} that the averaging must be undertaken after the propagator for each element of the distribution is raised to the $n^{\mathrm{th}}$ power. Since we have approximated the input density matrix of the system as being independent of this distribution, the superoperator is completely positive and has a Kraus representation \cite{Kraus:71a}:

\begin{equation}
  \rho_{\mathrm{out}} = \sum_{k=0}^{3}{A_{k} \rho_{\mathrm{in}} A_{k}^\dag},
\end{equation}
where the Kraus operators are given by

\begin{equation}
  A_{k} = \sqrt{p_{k}} U_{k},
  \label{eq:krausops}
\end{equation}
such that a portion of the sample given by the probability $p_{k}$ experiences evolution under the action of the unitary propagator $U_{k}$. Since we also require that the map be trace-preserving, the Kraus operators must satisfy the condition

\begin{equation}
  \sum{A_{k} A_{k}^\dag} = \mathrm{I}.
\end{equation}

The form of the Kraus operators given in eq. \ref{eq:krausops} is a result of the space of all completely positive, trace-preserving (CPTP) maps for a single spin being convex with unitary extrema. This means that any Kraus operator that is non-unitary may be written as a linear sum of unitary Kraus operators. Thus, any non-unitary Kraus representation of a CPTP map may be transformed into a linear combination of unitary operators. There are an infinite number of Kraus representations for a particular map, but only one maximizes the linear independence of the operators. This unique representation is determined by the eigenvectors and eigenvalues of the Choi matrix \cite{Choi:75a,Havel:03a}. 

A map corresponding to unitary evolution will yield only one non-zero eigenvalue and, as a result, be fully described by a single Kraus operator with unit probability. As the CPMG superoperator will not necessarily be unitary, the Kraus representation will generally consist of four operators with varying probabilities. Kraus representation as a Pauli channel yields considerable insight into the multi-pulse dynamics that occur during a CPMG sequence.

\subsection{Pauli Channel Model}
\label{subsec:paulichannel}
A Pauli channel is one of a class of quantum channels - a concept from the field of quantum information theory describing the nature of a transmission line between quantum mechanical entities \cite{NielsenChuang:00a}. In the context of the CPMG sequence, the quantum channel is the sequence of refocusing pulses and delays taken together with incoherent noise given by magnetic field inhomogeneities. The quantum mechanical entities to be connected are the initial and final density matrices representing the spin state after $n$ applications of the CPMG cycle. 

A Pauli channel, $C_{\mathrm{p}}$, of a CPTP map acting on an uncorrelated input state, $\rho_{\mathrm{in}}$, for an ensemble of spins-$\frac{1}{2}$ is a special case of a Kraus representation and is defined by

\begin{equation}
  \rho_{\mathrm{out}} = C_{\mathrm{p}}(\rho_{\mathrm{in}}) = \sum^{3}_{i=0}{p_{i} \sigma_{i} \rho_{\mathrm{in}} \sigma_{i}^\dag},
\label{eq:paulichanneldef}
\end{equation}
where $\sigma_{i} = \{I,\sigma_{x},\sigma_{y},\sigma_{z}\}$ and $p_{i}$ denotes the probability of the system undergoing evolution under the action of the $i^{\mathrm{th}}$ operator. For example, spin-spin (T$_2$) relaxation during a CPMG sequence with ideal pulses may be modeled as a Pauli channel dephasing process about the longitudinal axis, corresponding to $\sigma_z$-noise. For this process only two Kraus operators with non-zero probability are needed:

\begin{equation}\begin{split}
  &A_0(t) = \sqrt{0.5 + 0.5e^{-t/\mathrm{T}_2}} \hspace{0.5em} \mathrm{I} \\ 
  &A_1(t) = A_2(t) = 0 \\
  &A_3(t) = \sqrt{0.5 - 0.5e^{-t/\mathrm{T}_2}} \hspace{0.5em} \sigma_z
\end{split}
\end{equation}
$A_0$ refers to the identity operation, I, occurring with probability $p_0 = 0.5 + 0.5e^{-t/\mathrm{T}_2}$, while $A_3$ refers to a transverse dephasing operation, $\sigma_z$, occuring with probability $p_3 = 0.5 - 0.5e^{-t/\mathrm{T}_2}$. It is clear from this model that, for short times, all input states will experience the identity operation and remain unchanged. However, as time progresses, the probability of the dephasing operation, $\sigma_z$, occuring will exponentially grow, leading to the expected decay of the transverse spin states. Note, also, that the $\sigma_z$ input state, corresponding to longitudinal magnetization, will remain unchanged for all time, as desired. 

Similarly, if considered as a pure dephasing channel and assuming imperfect y-axis refocusing pulses, a Pauli channel representation of the CPMG sequence in the absence of relaxation will have two operations with non-zero probability, $\sigma_{y}$ and I. The $\sigma_y$ dephasing operator will cause the $\sigma_x$ and $\sigma_z$ input states to decay, while the $\sigma_y$ input state will remain unchanged. In general, though, CPMG dynamics are not pure dephasing and to properly model the sequence as a Pauli channel requires consideration of all four possible Kraus operators. It is evident, then, that a successful CPMG sequence with y-axis refocusing pulses must have asymptotically small and bounded probabilities for the $\sigma_x$ and $\sigma_z$ operators such that the $\sigma_y$ input state does not decay.

\begin{figure*}[!bt]\small
  \begin{center}
    \subfigure[Non-symmetrized OCT Pulse]{\includegraphics[scale=1,keepaspectratio]{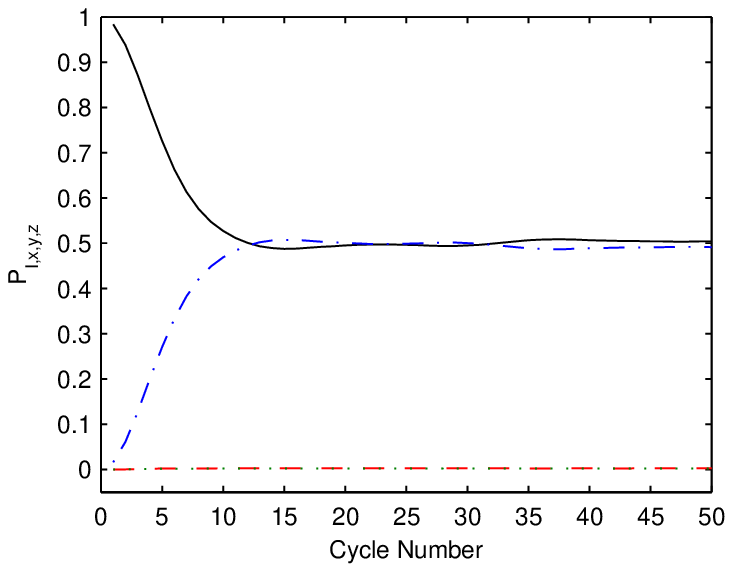}}
    \subfigure[Symmetrized OCT Pulse]{\includegraphics[scale=1,keepaspectratio]{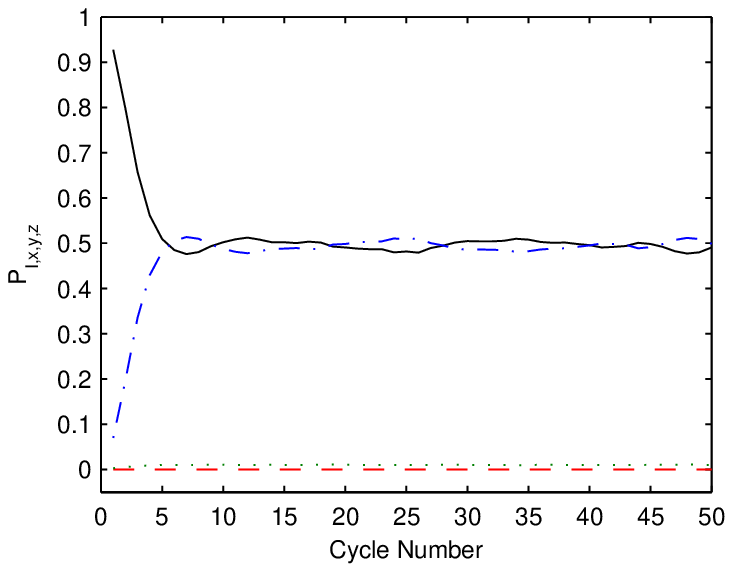}}
    \subfigure[Hard Pulse]{\includegraphics[scale=1,keepaspectratio]{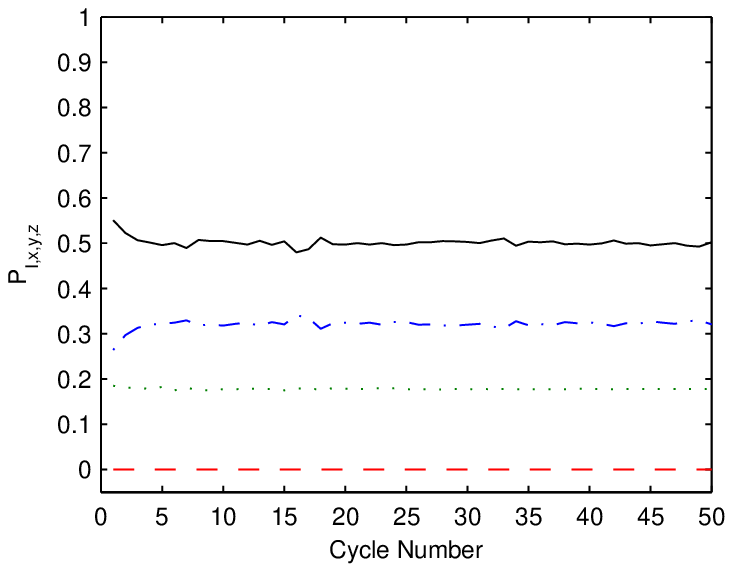}}
    \subfigure[Chirp Pulse]{\includegraphics[scale=1,keepaspectratio]{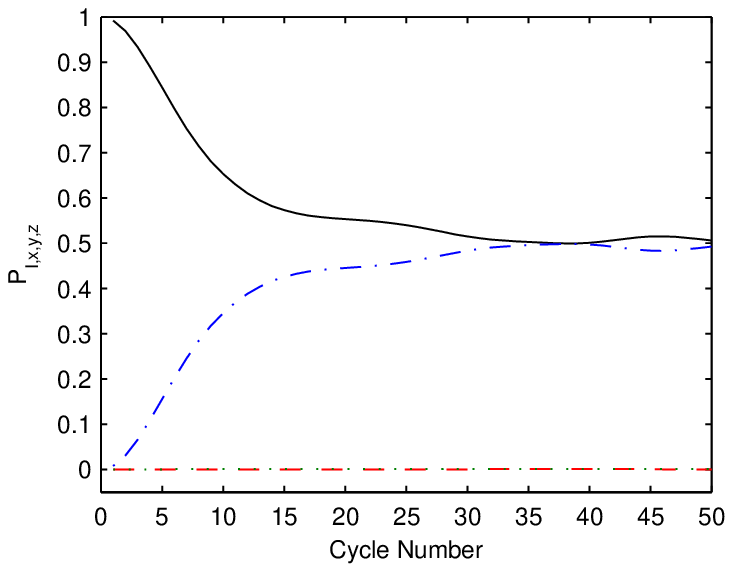}}
  \end{center}
    \caption{{\small [color online and in print] Pauli channel representations for CPMG dynamics using four different refocusing pulses. Probabilities of the identity (black solid), $\sigma_{x}$ (red dashed), $\sigma_{y}$ (blue dash-dotted), and $\sigma_{z}$ (green dotted) operations are shown as a function of the number of cycle applications. The sequences were simulated over $\left| \Delta\omega \right| \leq$ 1.6 $A_{\mathrm{max}}$ and $\omega_1 = 0.9 - 1.1$ with time between refocusing pulses of $2\tau = 2$ ms. Similar behavior is seen when the value of $\tau$ is varied. Representation as a Pauli channel allows us to accurately compare the influence of cumulative errors associated with each pulse. The dephasing rate constant, T$_{\mathrm{2,pulse}}$, was taken as the 1/e point in the decay of the identity probability.}}
    \label{fig:paulichannel}
\end{figure*}

The results of a Pauli channel fit for a CPMG sequence using our OCT refocusing pulse optimized over $\left| \Delta\omega \right| \leq$ 1.6 $A_{\mathrm{max}}$ and $\omega_1 = 0.9 - 1.1$ is shown in fig. \ref{fig:paulichannel}a. Based on a comparison of the fitted superoperators to the superoperators calculated directly from eq. \ref{eq:pobasissup}, the Pauli channel model captures over 99\% of the action of the map. It is evident that the decay and growth of the I and $\sigma_y$ operations are not entirely exponential, reinforcing the notion that the CPMG sequence cannot be fully represented as a pure dephasing channel. In fact, there are three main elements to the dynamics: An immediate loss of visibility due to the fringes of the distribution that are not compensated by the $\pi$-pulse, a fast loss of visibility due to initial oscillations that lead on large $n$ to damping, and dephasing that preserves only a single orthogonal input state.

\begin{figure}[!hbt]
  \begin{center}
    \subfigure[CPMG Sequence]{\includegraphics[scale=1,keepaspectratio]{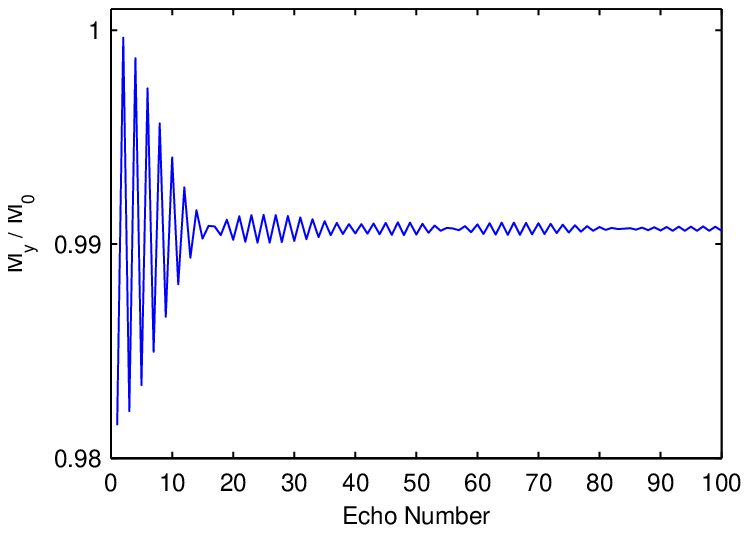}}
    \subfigure[CP Sequence]{\includegraphics[scale=1,keepaspectratio]{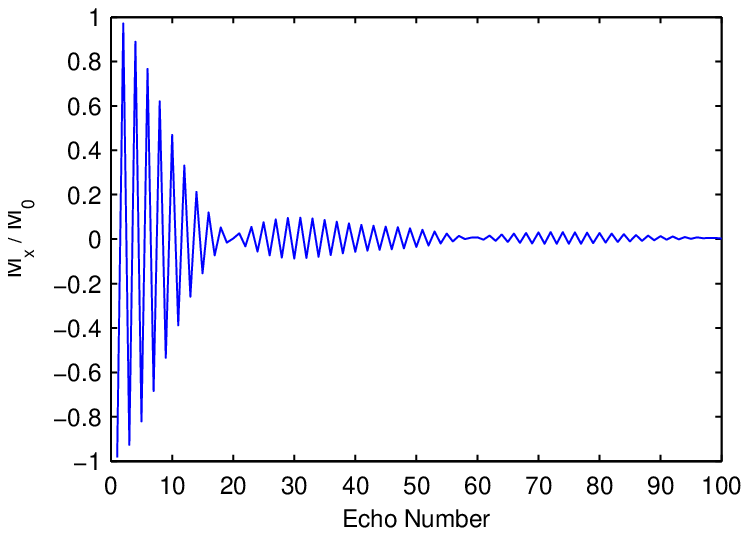}}
  \end{center}
    \caption{{\small [color online and in print] Full simulated dynamics for a CPMG sequence and CP sequence using our OCT refocusing pulse optimized and simulated over $\left| \Delta\omega \right| \leq$ 1.6 $A_{\mathrm{max}}$ and a range of RF amplitudes of $\omega_1 = 0.9 - 1.1$. The time between refocusing pulses was set to $2\tau = 2$ ms. The immediate and fast loss of visibility is shown by the initial and asymptotic magnetization being less than unity. The decay of the transients is due to the dephasing portion of the dynamics. Representation of the dynamics as a Pauli channel captures over 99$\%$ of the action of the map, describing both the loss of visibility and pulse-induced dephasing.}}
    \label{fig:cpmgvscp}
\end{figure}

The complete simulated dynamics of a CPMG sequence and CP sequence using our OCT refocusing pulse is shown in fig. \ref{fig:cpmgvscp}. Note that the complete CP(MG) dynamics are calculated for every echo while the Pauli channel representation is calculated over the cycle (2 echoes). The decay envelope of the CP sequence ($\rho_{\mathrm{in}} = \sigma_x$) is captured almost entirely by the decay and growth of the probabilities of the Pauli channel identity and $\sigma_y$ operations, respectively. It is the $\sigma_y$ and $\sigma_z$ operations that represent the dephasing part of the dynamics for a CP sequence. As the probability of the $\sigma_y$ operation occuring is not asymptotically small and bounded, the magnetization retained by the CP sequence is nearly zero for large numbers of echoes. This contrasts with the envelope of the CPMG sequence ($\rho_{\mathrm{in}} = \sigma_y$), where the dephasing operators are $\sigma_x$ and $\sigma_z$. As the probabilities of these operators are asymptotically small and bounded, the magnetization retained by the CPMG sequence is always close to unity. 

A method for quantitatively evaluating the cumulative pulse errors that occur during a CPMG sequence can now be formulated. We may assign a model based on a Pauli channel that captures the three elements of the CPMG dynamics just mentioned: 

\begin{equation}\begin{split}
   &A_0(n) = \sqrt{c_I + 0.5e^{-n t_{\mathrm{c}} /\mathrm{T}_{\mathrm{2,pulse}}}} \hspace{0.5em} \mathrm{I} \\
   &A_1(n) = \sqrt{c_x} \hspace{0.5em} \sigma_x \\
   &A_2(n) = \sqrt{c_y - 0.5e^{-n t_{\mathrm{c}} /\mathrm{T}_{\mathrm{2,pulse}}}} \hspace{0.5em} \sigma_y \\
   &A_3(n) = \sqrt{c_z} \hspace{0.5em} \sigma_z
\end{split}
\end{equation}
In this model, T$_{\mathrm{2,pulse}}$ is a time constant which represents the dephasing that occurs due to the pulse errors and $t_c$ is the time between every 2 echoes (the cycle time). For $n << \mathrm{T}_{\mathrm{2,pulse}}/t_{\mathrm{c}}$ the CPMG and CP sequences perform roughly the same as an identity operation, while for $n >> \mathrm{T}_{\mathrm{2,pulse}}/t_{\mathrm{c}}$ the $\sigma_y$ operator becomes significant, causing the CP sequence to retain no initial magnetization. Note that the probabilities are now a function of cycle number to reflect the distinction that T$_2$ is independent of echo spacing, while T$_{\mathrm{2,pulse}}$ is not. The constants, $c_{I,x,y,z}$, determine the asymptotic behavior of the channel and represent loss of visibility. At the cost of losing information about the initial magnetization oscillations present in any CPMG sequence, the constants $c_x$ and $c_z$ are taken to be independent of $n$, such that the overall loss of visibility due to pulse errors is still retained. The asymptotically retained magnetization for a CPMG sequence is given by $M_{\infty} = c_I + c_y - (c_x + c_z)$ and the rate at which a CP sequence would lose signal is given by T$_{\mathrm{2,pulse}}$. $M_{\infty}$ and T$_{\mathrm{2,pulse}}/t_{\mathrm{c}}$ are two numbers that allow us to completely and compactly characterize refocusing pulses for use in multiple refocusing sequences. The ideal refocusing pulse has T$_{\mathrm{2,pulse}}/t_{\mathrm{c}} \rightarrow \infty$ and $M_{\infty} \rightarrow 1$. 

The result of a Pauli channel fit for CPMG sequences using the four pulses previously compared in section \ref{subsec:comparison} is displayed in fig. \ref{fig:paulichannel}. All pulses were restricted to a maximum amplitude of $A_{\mathrm{max}} / 2\pi = 5$ kHz and were simulated over $\left| \Delta\omega \right| \leq$ 1.6 $A_{\mathrm{max}}$ and $\pm$10$\%$ RFI ($\omega_1 = 0.9 - 1.1$). The difference in refocusing ability for each pulse is clearly distinguishable (see table \ref{tab:pulsecompare}). 

\begin{table}[!h]
\begin{center}
\begin{tabular}{c|c|c}
Pulse & T$_{\mathrm{2,pulse}}/t_{\mathrm{c}}$ & $M_\infty$ \\[0.5ex]
\hline
Chirp & 9 cycles & 0.997 \\
Non-symmetrized OCT & 6 cycles & 0.991 \\
Symmetrized OCT & 3 cycles & 0.981 \\
Hard & 1 cycle & 0.646
\end{tabular} 
\end{center}
\caption{{\small Dephasing rate and asymptotically retained magnetization for CPMG sequences using four different types of refocusing pulses.}}
\label{tab:pulsecompare}
\end{table}


\subsection{Validity of Pauli Channel}
\label{subsec:validity}
A Pauli channel model is an excellent description of CPMG dynamics due to both the refocusing being about a single cartesian coordinate and the averaging being complete enough over the distribution in question to suppress the off-diagonal components of the superoperator (eq. \ref{eq:pobasissup}). A superoperator that may be accurately represented as a Pauli channel must be diagonal in the Pauli basis. This requirement is derived by noting that a Pauli channel (as defined by eq. \ref{eq:paulichanneldef}) does not mix the components of the input state. For example, for each orthogonal input state ($\sigma_x$, $\sigma_y$, and $\sigma_z$) the output from the channel will remain either $\sigma_x$, $\sigma_y$, or $\sigma_z$, but with a scaling factor. In general, though, for an arbitrary distribution, $P(\Delta\omega,\omega_1)$, and a refocusing pulse about an arbitrary axis, the superoperator describing the dynamics in the Pauli basis will contain off-diagonal components. This can be seen by considering that for each element of the distribution, the cycle propagator is of the general form

\begin{equation}
  U_{\mathrm{cycle}}(\Delta\omega,\omega_{1}) = e^{-i \frac{\theta(\Delta\omega,\omega_{1})}{2} \, \widehat{r}(\Delta\omega,\omega_{1})\cdot \vec{\sigma}}.
\label{eq:generalcycle}
\end{equation}
The resulting density matrix after $n$ applications of the cycle, for an uncorrelated initial state $\rho_{\mathrm{in}}$, is then given by

\begin{equation}\begin{split}
  \rho^{(n)}_{\mathrm{out}}& = \int P(\Delta\omega,\omega_{1})\bigg[\cos^{2}\frac{n\theta}{2} (\mathrm{I} \,  \rho_{\mathrm{in}} \, \mathrm{I}) + \\& i\sin\frac{n\theta}{2}\cos\frac{n\theta}{2} (\mathrm{I} \, \rho_{\mathrm{in}} \, \widehat{r}\cdot\vec{\sigma}) - \\& i\sin\frac{n\theta}{2}\cos\frac{n\theta}{2} (\widehat{r}\cdot\vec{\sigma} \, \rho_{\mathrm{in}} \, \mathrm{I}) + \\& \sin^{2}\frac{n\theta}{2} (\widehat{r}\cdot\vec{\sigma} \, \rho_{\mathrm{in}} \, \widehat{r}\cdot\vec{\sigma})\bigg] \, \mathrm{d}\Delta\omega \, \mathrm{d}\omega_1.
\end{split}
\label{eq:expandedprop}
\end{equation}
This equation is an expansion of eq. (\ref{eq:generalcycle}) inserted into eq. (\ref{eq:pobasissup}). The dependence of $\theta$ and $\widehat{r}$ on $\Delta\omega$ and $\omega_{1}$ has been dropped for clarity and the argument $n\theta$ is taken to be modulo $2\pi$. As noted in Ref. \cite{HurlimannGriffin:00a} using analysis in SO(3), if after a certain number of cycles the range $\theta = \left[0,2\pi\right]$ is sampled uniformly over $\Delta\omega$ and $\omega_{1}$, eq. (\ref{eq:expandedprop}) reduces to

\begin{equation}\begin{split}
  \rho_{\mathrm{out}} =& \int P(\Delta\omega,\omega_1) \bigg[ \frac{1}{2} \left( \mathrm{I} \, \rho_{\mathrm{in}} \, \mathrm{I} \right) + \\& \frac{1}{2} \left( \widehat{r}\cdot\vec{\sigma} \right) \rho_{\mathrm{in}} \left( \widehat{r}\cdot\vec{\sigma} \right) \bigg] \, \mathrm{d}\Delta\omega \, \mathrm{d}\omega_1.
\end{split}\end{equation}
The integrand may be expanded as

\begin{equation}\label{eq:bigavg}\begin{split}
  \frac{1}{2} & \left( \mathrm{I} \, \rho_{\mathrm{in}} \, \mathrm{I} \right) + \frac{1}{2}\bigg[r_{x}^{2} \left( \sigma_{x}\rho_{\mathrm{in}}\sigma_{x} \right) + r_{y}^{2} \left( \sigma_{y}\rho_{\mathrm{in}}\sigma_{y} \right) + \\& r_{z}^{2} \left( \sigma_{z}\rho_{\mathrm{in}}\sigma_{z} \right) + r_{x}r_{y} \left( \sigma_{x}\rho_{\mathrm{in}}\sigma_{y} \right) + \\& r_{x}r_{z} \left( \sigma_{x}\rho_{\mathrm{in}}\sigma_{z} \right)  + r_{y}r_{x} \left( \sigma_{y}\rho_{\mathrm{in}}\sigma_{x} \right) + \\& r_{y}r_{z} \left( \sigma_{y}\rho_{\mathrm{in}}\sigma_{z} \right) + r_{z}r_{x} \left( \sigma_{z}\rho_{\mathrm{in}}\sigma_{x} \right) + \\& r_{z}r_{y} \left( \sigma_{z}\rho_{\mathrm{in}}\sigma_{y} \right) \bigg].
\end{split}\end{equation} 
The same result may be arrived at by computing the eigenvalues of the Choi matrix, as outlined in \cite{Havel:03a}.

The averaging of the off-diagonal elements of the superoperator in the product operator basis - elements of the form $\sigma_{i}\rho_{\mathrm{in}}\sigma_{j}$, where $i\neq j$ - is dependent upon the distribution and refocusing pulse in question. The inhomogeneity we consider here is symmetric about the on-resonance Larmor precession frequency such that the average of $r_z$ over $\Delta\omega$ and $\omega_1$ will tend toward zero. Additionally, field inhomogeneities tend to tip the effective rotation axis of a refocusing pulse out of the transverse plane while minimally affecting the orientation in the transverse plane. Since we are considering a y-axis pulse, $r_x$ will tend to be small for all values of $\Delta\omega$ and $\omega_1$. Thus, the only terms to be significant in eq. (\ref{eq:bigavg}) will be $r_{x,y,z}^{2}$ such that the dynamics is given by a Pauli channel to a high degree of accuracy. 

\begin{figure*}[!bt]\small
  \begin{center}
    \subfigure[OCT $r_{y}$]{\includegraphics[scale=1]{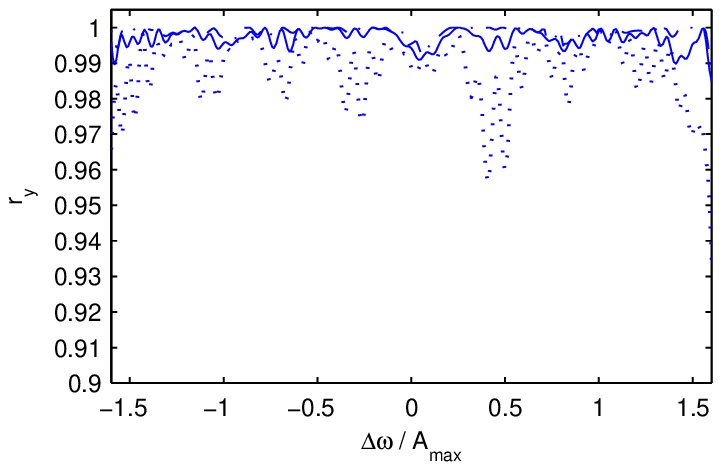}}
    \subfigure[Hard $r_{y}$]{\includegraphics[scale=1]{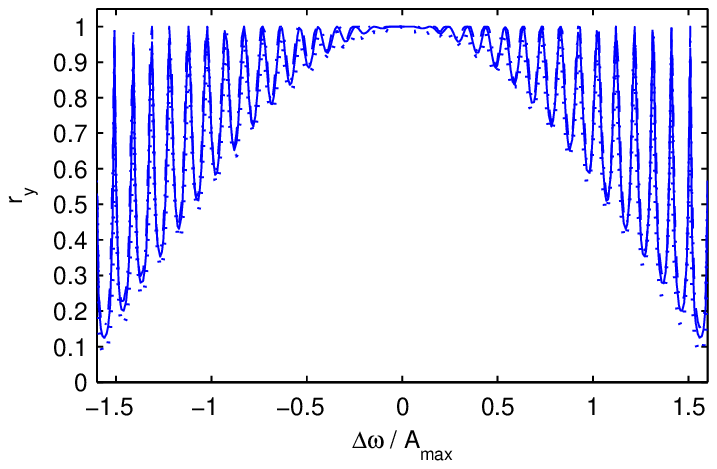}}
    \subfigure[OCT 1 Cycle]{\includegraphics[scale=1]{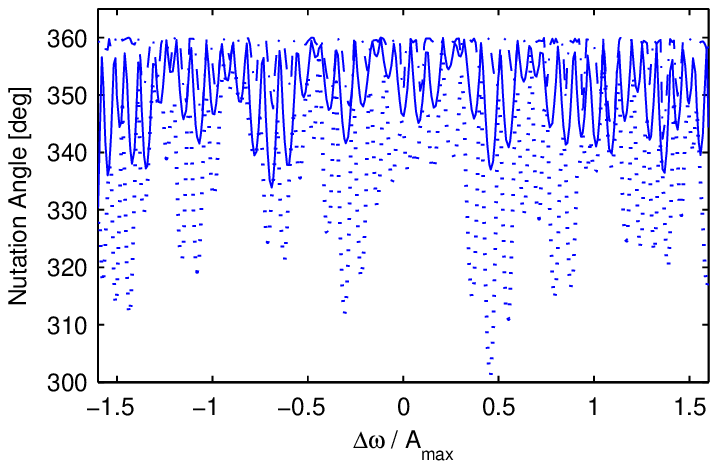}}
    \subfigure[Hard 1 Cycle]{\includegraphics[scale=1]{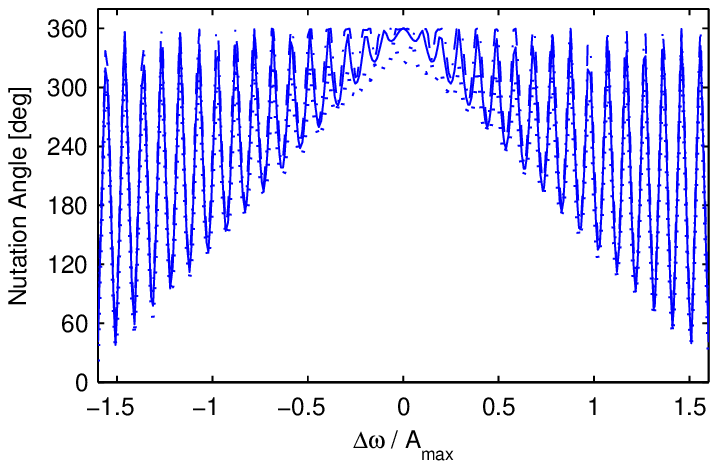}}
    \subfigure[OCT 50 Cycles]{\includegraphics[scale=1]{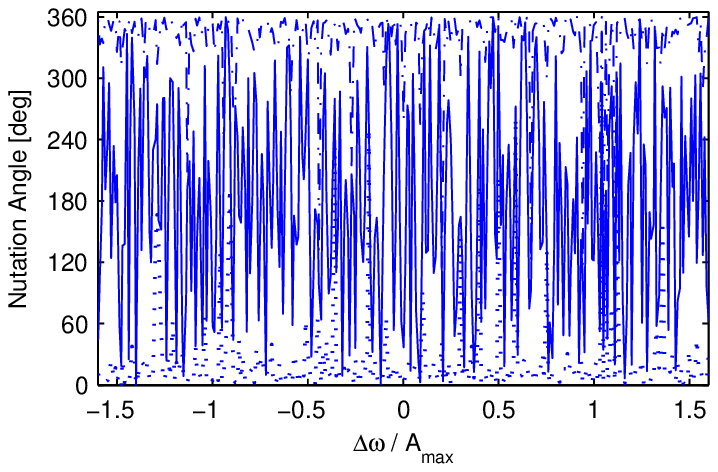}}
    \subfigure[Hard 50 Cycles]{\includegraphics[scale=1]{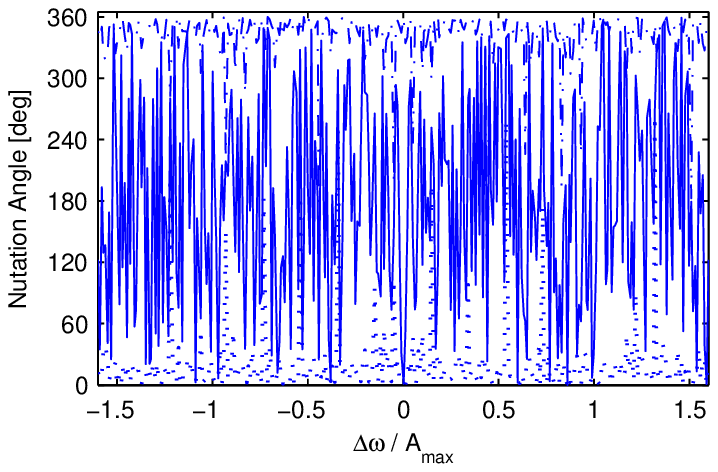}}
  \end{center}
  \caption{{\small [color online and in print] Decomposition of the propagator characterizing a cycle [$\tau-\pi_y - 2 \tau - \pi_y - \tau$] with OCT pulses (left) and standard hard pulses (right) versus resonance offset. The top panels show the projections of the net rotation axis onto the $\widehat{y}$ axis, the middle panels the nutation angle for a single cycle, and the bottom panels the effective nutation angle for 50 cycles. In each panel, the solid line is the response for uniform RF ($\omega_1$ = 1), while the dashed (dotted) lines indicate the maximum (minimum) over the range of $\omega_1$ = 0.9 - 1.1. The OCT pulse was optimized over $\left| \Delta\omega \right| \leq$ 1.6 $A_{\mathrm{max}}$ with a uniform range of RF of $\omega_1$ = 0.9 - 1.1. For this calculation $\tau$ was set to 1 ms. Note that in the top two panels, the scales for the OCT and hard pulses are different.}}
  \label{fig:cycledecomp}
\end{figure*}

In the case of our OCT CPMG cycle, as shown in fig. \ref{fig:cycledecomp}, $\widehat{r}\cdot\vec{\sigma}$ is very nearly $\sigma_{y}$ for all $\Delta\omega$ and $\omega_{1}$ and the variation in $\theta$ across $\Delta\omega$ and $\omega_{1}$ is such that after roughly 30 cycle applications the averaging is complete enough that the map is given almost entirely by I$\rho_{\mathrm{in}}$I and $\sigma_{y}\rho_{\mathrm{in}}\sigma_{y}$ with equal probabilities. For the field inhomogeneities considered here ($\left| \Delta\omega/2\pi \right| \leq$ 8 kHz and $\omega_1 = 0.9 - 1.1$), the model is an excellent description of the dynamics, with a trace overlap of greater than 0.9999 between the fitted and simulated superoperators for all cycles. Even for the case of hard pulses where the variation of $r_{y}$ and $\theta$ is large over a single cycle (see fig. \ref{fig:cycledecomp}), the trace overlap is still greater than 0.999 for all cycles.

\section{Conclusions}
\label{sec:conclusions}
We have applied optimal control techniques to find general refocusing pulses that act as a universal $\pi)_y$ rotation over a wide range of resonance offset frequencies and RF amplitudes. Such pulses can significantly extend the range over which the CPMG sequence retains its intrinsic robust nature. An iterative optimization procedure was used to systematically expand the target bandwidth while keeping the total duration and maximum RF amplitude of the pulses fixed. We specifically considered pulses that are 1 ms long, with limited instantaeous RF power of $\gamma B_{\mathrm{1,max}}/2\pi = 5$ kHz. This specific problem is directly relevant to noise suppression in QIP applications and for decoupling applications. It is also an important first step in establishing the framework for future investigations for other applications, including finding optimized OCT pulses that maximize the SNR of the CPMG sequence in grossly inhomogeneous fields.

A significant advantage of using OCT techniques is the flexibility of the method. It allows addressing unique requirements and constraints in individual experimental situations. This is accomplished by redefining the measures of pulse performance. In general, one pulse will not ideally suit the needs of all applications. The versatility of OCT lies in the ease with which various performance functionals and constraints may be substituted into the optimization procedure without requiring a change to the general optimization methods.

The performance of our pulses with respect to the CPMG criteria demonstrates a significant reduction in the variation of the net nutation angle and direction of the rotation axis over the optimization range when compared to a hard pulse. Although the variation is not fully eliminated, we were able to find a pulse which refocuses over 99$\%$ of the initial magnetization over a range of frequency offsets 4 times the maximum amplitude of the RF power ($\pm$10 kHz) for uniform RF. In the case when the RF has an amplitude distribution of $\pm$10$\%$, we were able to find a pulse that refocuses over 98$\%$ of the initial magnetization over a frequency range of 3.2 times the maximum RF amplitude ($\pm$8 kHz). 

The performance of our pulses also demonstrates improvement over any previously published refocusing pulses of similar time and maximum RF amplitude. Comparable performance to our directly optimized pulse was only found when we symmetrized a previously published broadband OCT excitation pulse to make it a universal $\pi$-pulse. Repeated application of the refocusing pulses does not induce any extra decay due to pulse errors. This enables the monitoring of the magnetization decay with a large number of echoes. The observed relaxation time is dominated by the $T_2$ of the sample, with some minor $T_1$ contamination during the pulses. Additionally, we have demonstrated that our OCT pulses are experimentally viable and perform in experiments as predicted by simulation. 

Superoperators were used to analyze the general spin dynamics for a pulse over a system with a distribution of $B_0$ and $B_1$ fields. As demonstrated, the spin dynamics are well described as a Pauli channel, which provides a compact form to predict the performance of multiple pulses, both for CP and CPMG sequences, and allows the performance of the refocusing pulses to be succinctly described by only two parameters: the asymptotic visibility, $M_\infty$, and the pulse-induced dephasing time, $T_{\mathrm{2,pulse}}$, of the sequence. Additionally, a recent discovery in quantum information theory states that any noise which may be described as a Pauli channel may be compensated by an appropriate logical encoding \cite{BlumeKohoutViola:08a}. This allows, through the use of encoding the logical state of a single spin into the physical state space of two spins, a CPMG sequence which treats all input states identically, yielding the identity channel for all times even in the presence of decoherence and dynamic pulse errors. 

Other future applications of our results and the methods described in this work include minimizing $T_1$ contamination during pulsing by limiting the allowed trajectories on the Bloch sphere, eliminating transients by designing pulses that dephase immediately due to pulse errors (i.e. $T_{\mathrm{2,pulse}} = 0$), increasing bandwidth by directly optimizing the rotation axis over a cycle, sculpting response of excitation pulses to directly match the rotation axis of the CPMG cycle for each element of the field distribution, systematically optimizing SNR by examining the maximum bandwidth achievable for a variety of pulse lengths, and designing pulses that minimize overall power consumption without significantly affecting bandwidth. However, as the optimal solutions, in both the symmetrized and our unsymmetrized versions (see supplementary material), seem to have RF at maximal amplitude for almost the entire pulse duration, optimizing for overall power consumption becomes a problem of once again maximizing bandwidth for different pulse lengths to achieve the best SNR. These applications are just a few of many novel ways in which optimal control techniques may be used to design more robust CPMG sequences. 

\section{Acknowledgments}
We thank Chandrasekhar Ramanathan and the NMR group at Schlumberger-Doll Research (SDR) for many useful discussions. This work was supported in part by the National Security Agency (NSA) under Army Research Office (ARO) contract number W911NF-05-1-0469. TWB also acknowledges and thanks SDR for partial support during the completion of this work. DGC would also like to acknowledge support from the Canadian Excellence Research Chairs (CERC) program. 
 
\bibliographystyle{elsarticle-num}
\bibliography{AMoreRobustCPMGSequence}

\end{document}